\documentclass[twocolumn,showpacs,preprintnumbers,amsmath,amssymb]{revtex4}

\usepackage{graphicx}
\usepackage{dcolumn}
\usepackage{amsfonts,amsmath,amssymb,bm}

\begin{document}

\title{Atomic self-interaction correction for molecules and solids}

\author{C.~D.~Pemmaraju$^1$, T.~Archer$^1$, D.~S\'anchez-Portal$^2$ and S.~Sanvito$^1$}
\affiliation{$^1$ School of Physics, Trinity College, Dublin 2, Ireland}
\affiliation{$^2$ Unidad de Fisica de Materiales, Centro Mixto CSIC-UPV/EHU and Donostia International Physics Center (DIPC), Paseo Manuel de Lardizabal 4, 20018 Donostia-San Sebasti\'an, Spain}

\date{\today}

\begin{abstract}
We present an atomic orbital based approximate scheme for self-interaction correction (SIC) to the local density approximation of 
density functional theory. The method, based on the idea of Filippetti and Spaldin [\cite{alessio} Phys. Rev. B {\bf 67}, 125109 (2003)],
is implemented in a code using localized numerical atomic orbital basis sets
and is now suitable for both molecules and extended solids. After deriving the fundamental equations
as a non-variational approximation of the self-consistent SIC theory, we present results for a wide range of molecules and
insulators. In particular, we investigate the effect of re-scaling the self-interaction
correction and we establish a link with the existing atomic-like corrective scheme LDA+$U$. We find that when
no re-scaling is applied, i.e. when we consider the entire atomic correction, 
the Kohn-Sham HOMO eigenvalue is a rather good approximation to the experimental
ionization potential for molecules. Similarly the HOMO eigenvalues of negatively charged molecules reproduce
closely the molecular affinities. In contrast a re-scaling of about 50\% is necessary to reproduce insulator bandgaps in solids, 
which otherwise are largely overestimated. The method therefore represents a Kohn-Sham based single-particle 
theory and offers good prospects for applications where the actual position of the Kohn-Sham eigenvalues is
important, such as quantum transport.
\end{abstract}


\maketitle

\section{Introduction} 

Density functional theory (DFT), in both its static \cite{DFT} and time-dependent
\cite{gross} forms, has become by far the most successful and widely used among all 
the electronic structure methods. The most obvious reason for this success is that it 
provides accurate predictions of numerous properties of atoms, inorganic molecules, bio-molecules, 
nanostructures and solids, thus serving different scientific communities. 

In addition DFT has a solid theoretical foundation. The Hohenberg-Kohn theorem \cite{DFT}
establishes the existence of a unique energy functional $E[\rho]$ of the electron density $\rho$ which 
alone is sufficient to determine the exact ground-state of a $N$-electron system. Although the energy 
functional itself is not known, several of its general properties can be demonstrated rigorously. These are 
crucial for constructing increasingly more predictive approximations to the functional and for addressing
the failures of such approximations\cite{perdew1}.

Finally, but no less important, the Kohn-Sham (KS) formulation of DFT \cite{KS} establishes a one to one
mapping of the intrinsically many-body problem onto a fictitious single-particle system and offers a
convenient way for minimizing $E[\rho]$. The degree of complexity of the Kohn-Sham (KS) problem 
depends on the approximation chosen for the density functional. In the case of the local density approximation 
(LDA) \cite{KS} the KS problem typically scales as $N^3$, where the scaling is dominated by the diagonalization 
algorithm.  However, clever choices with regards to basis sets and sophisticated numerical methods make order-$N$ 
scaling a reality \cite{orderN,orderN2}. 

The energy functional $E[\rho^\uparrow,\rho^\downarrow]$ ($\rho^\sigma$, $\sigma=\uparrow, \downarrow$ is the spin
density, $\rho=\sum_\sigma\rho^\sigma$) can be written as 
\begin{equation}
E[\rho^\uparrow,\rho^\downarrow]=T_\mathrm{S}[\rho]+\int\mathrm{d}^3{\bf r}\:\rho({\bf r})v({\bf r})+U[\rho]+E_\mathrm{xc}[\rho^\uparrow,\rho^\downarrow]\;,
\label{enfunc}
\end{equation}
where $T_\mathrm{S}$ is the kinetic energy of the non-interacting system, $v({\bf r})$ the external
potential, $U$ the Hartree electrostatic energy and $E_\mathrm{xc}$ the exchange and
correlation (XC) energy. This last term is unknown and must be approximated. 
The construction of an approximated functional follows two 
strategies: empirical and ``constraint satisfaction''.

Empirical XC functionals usually violate some of the constraints imposed by 
exact DFT, and rely on parameterizations obtained by fitting representative data. One includes in
this category, XC functionals which borrow some functional dependence from other theories. This is
for instance the case of the celebrated LDA+$U$ scheme \cite{AnisimovZaanen,AnisimovLongPaper}, where
the Hubbard-$U$ energy takes the place of the LDA energy for certain ``strongly correlated'' atomic orbitals 
(typically $d$ and $f$ shells). The method however depends on the knowledge of the Coulomb and exchange 
parameters $U$ and $J$, which vary from material to material, and can also be different for the same 
ion in different chemical environments \cite{degir,gosia}.

In contrast the construction based on ``constraint satisfaction'' proceeds by developing increasingly 
more sophisticated functionals, which nevertheless satisfy most of the properties of exact DFT \cite{XChole}. 
It was argued that this method constructs a ``Jacob's ladder'' \cite{jacob}, where functionals are assigned 
to different rungs depending on the number of ingredients they include. Thus the LDA, which depends only 
on the spin-densities is on the first rung, the generalized gradient approximation (GGA) \cite{PBE}, which depends 
also on $\boldsymbol{\nabla}\rho^\sigma$, is on the second rung, the so-called meta-GGA functionals \cite{meta}, 
which in addition to $\rho^\sigma$ and $\boldsymbol{\nabla}\rho^\sigma$ depend on either the Laplacian 
$\boldsymbol{\nabla}^2\rho^\sigma$ or the orbital kinetic energy density, are on the third rung and so on.
The higher its position on the ladder the more accurate a functional becomes, 
but at the price of increasing computational overheads. Therefore its worth investigating 
corrections to the functionals of the lower rungs, which preserve most of the fundamental properties of DFT 
and do not generate massive additional numerical overheads.

One of the fundamental problems intrinsic to the semi-local functionals of the first three rungs is the presence of
self-interaction (SI) \cite{PZSIC}. This is the spurious interaction of an electron in a given KS orbital with the
Hartree and XC potential generated by itself. Such an interaction is not present in the Hartree-Fock method,
where the Coulomb self-interaction of occupied orbitals is cancelled exactly by the non-local exchange.
However when using semi-local functionals such a cancellation is not complete and the 
rigorous condition for KS-DFT
\begin{equation}
U[\rho_n^\sigma]+E_\mathrm{xc}[\rho_n^\sigma,0]=0
\end{equation}
for the orbital density $\rho_n^\sigma=|\psi_n^\sigma|^2$ of the fully occupied KS orbital $\psi_n^\sigma$ 
is not satisfied. A direct consequence of the self-interaction in LDA/GGA is that the KS potential becomes too
repulsive and exhibits an incorrect asymptotic behavior \cite{PZSIC}. 

This ``schizophrenic''
(self-interacting) nature of semi-local KS potentials generates a number of failures in describing 
elementary properties of atoms, molecules and solids. Negatively charged ions (H$^-$, O$^-$,
F$^-$) and molecules are predicted to be unstable by LDA \cite{neions} and transition metal oxides
are described as small-gap Mott-Hubbard antiferromagnets (MnO, NiO) \cite{motthubbard} 
or even as ferromagnetic metals (FeO, CoO) \cite{motthubbard} instead of charge-transfer insulators.
Moreover the KS highest occupied molecular orbital (HOMO), the only KS eigenvalue that can be
rigorously associated to a single particle energy \cite{janak,PPLB,PL97}, 
is usually nowhere near the actual ionization potential \cite{PZSIC}.

Finally XC functionals affected by SI do not present a derivative
discontinuity as a function of the occupation \cite{janak,PPLB}. Semi-local functionals in fact continuously 
connect the orbital levels of systems of different integer occupation. This means for instance that when adding an extra 
electron to an open shell $N$-electron system the KS potential does not jump discontinuously by $I_\mathrm{N}-
A_\mathrm{N}$ where $I_\mathrm{N}$ and $A_\mathrm{N}$ are respectively the ionization potential and
the electron affinity for the $N$-electron system. This serious drawback is responsible for the incorrect
dissociation of heteronuclear molecules into charged ions \cite{Pb85} and for the metallic conductance
of insulating molecules \cite{cormac}.

The problem of removing the SI from a semi-local potential was acknowledged a long
time ago when Fermi and Amaldi proposed a first crude correction \cite{FermiAmaldi}. However the
modern theory of self-interaction corrections (SIC) in DFT is due to the original work of Perdew and Zunger 
from almost three decades ago \cite{PZSIC}. Their idea consists in removing directly the 
self-Hartree and self-XC energy of all the occupied KS orbitals 
from the LDA XC functional (the same argument is valid for other semi-local functionals), 
thus defining the SIC functional as
\begin{equation}
E_\mathrm{xc}^\mathrm{SIC}[\rho^\uparrow,\rho^\downarrow]=E_\mathrm{xc}^\mathrm{LDA}[\rho^\uparrow,\rho^\downarrow]-
\sum_{n\sigma}^\mathrm{occupied}(U[\rho_n^\sigma]+E_\mathrm{xc}^\mathrm{LDA}[\rho_n^\sigma,0])\;.
\label{EFSIC}
\end{equation}

Although apparently simple, the SIC method is more involved than standard KS DFT.
The theory is still a density functional one, i.e. it satisfies the Hohenberg-Kohn theorem, however it does
not fit into the Kohn-Sham scheme, since the one-particle potential is orbital-dependent. This means
that one cannot define a kinetic energy functional independently from the choice
of $E_\mathrm{xc}$ \cite{PZSIC}. Two immediate consequences are that the $\psi_n^\sigma$ are not
orthogonal, and that the orbital-dependent potential can break the symmetry of the system.
This last aspect is particularly important for solids since one has to give up the Bloch representation.

In this paper we explore an approximate method for SIC to the LDA, 
which has the benefit of preserving the local nature of the LDA potential, and therefore maintains 
all of the system's symmetries. We have followed in the footsteps of Filippetti and Spaldin \cite{alessio},
who extended the original idea of Vogel and co-workers \cite{voeg1,voeg2,voeg3} of considering
only the atomic contributions to the SIC. 
We have implemented such a scheme into the localized atomic orbital code {\it Siesta} \cite{siesta} and applied 
it to a vast range of molecules and solids. In particular we have investigated in detail how the scheme 
performs as a single-particle theory, and how the SIC should be rescaled in different chemical environments.


\section{Review of existing methods}

The direct subtraction proposed by Perdew and Zunger is
the foundation of the modern SIC method. However the minimization of the SIC functional
(\ref{EFSIC}) is not trivial, in particular for extended systems. The main problem is that
$E_\mathrm{xc}$ itself depends on the KS orbitals. Thus it does not fit into the standard KS scheme and 
a more complicated minimization procedure is needed. 

Following the minimization strategy proposed by Levy \cite{levy}, which prescribes to minimize the functional
first with respect to the KS orbitals $\psi_n^\sigma$ and then with respect to the occupation numbers $p_n^\sigma$,
Perdew and Zunger derived a set of single-particle equations
\begin{equation}
\left[-\frac{1}{2}\boldsymbol{\nabla}^2+v_{\mathrm{eff},n}^\sigma({\bf r})\right]\psi_n^\sigma=\epsilon_{n}^{\sigma,\mathrm{SIC}}
\psi_n^\sigma\;,
\label{KSSIC}
\end{equation}
where the effective single-particle potential $v_{\mathrm{eff},n}^\sigma({\bf r})$ is defined as
\begin{eqnarray}
v_{\mathrm{eff},n}^\sigma({\bf r})=v({\bf r})+u([\rho]; {\bf r})+v_\mathrm{xc}^{\sigma,\mathrm{LDA}}([\rho^\uparrow,\rho^\downarrow]; {\bf r})-\nonumber \\
-u([\rho_n]; {\bf r})-v_\mathrm{xc}^{\sigma,\mathrm{LDA}}([\rho_n^\uparrow,0]; {\bf r})
\;,
\label{KSSIC-pot}
\end{eqnarray}
and 
\begin{eqnarray}
u([\rho]; {\bf r})=\int\mathrm{d}^3{\bf r}^\prime\frac{\rho({\bf r}^\prime)}{|{\bf r}-{\bf r}^\prime|}\;, \\
v_\mathrm{xc}^{\sigma,\mathrm{LDA}}([\rho^\uparrow,\rho^\downarrow]; {\bf r})=\frac{\delta}{\delta \rho^\sigma({\bf r})}
E_\mathrm{xc}^\mathrm{LDA}[\rho^\uparrow,\rho^\downarrow]\;.
\end{eqnarray}

These are solved in the standard KS way for atoms, with good results in terms of quasi-particle
energies \cite{PZSIC}. In this particular case the KS orbitals $\psi_n^\sigma$ show only small deviations 
from orthogonality, which is re-imposed with a standard Schmidt orthogonalization. 

The problem of the non-orthogonality of the KS orbitals can be easily solved by imposing the 
orthogonality condition as a constraint to the energy functional, thus obtaining the following
single-particle equation
\begin{equation}
\left[-\frac{1}{2}\boldsymbol{\nabla}^2+v_{\mathrm{eff},n}^\sigma({\bf r})\right]\psi_n^\sigma
=\sum_m\epsilon_{nm}^{\sigma,\mathrm{SIC}}\psi_m^\sigma\;.
\label{KSSIC-orth}
\end{equation}
Even in this case where orthogonality is imposed, two major problems remain: the orbitals minimizing 
the energy functional are not KS-type and in general do not satisfy the system's symmetries.

If one insists in minimizing the energy functional in a KS fashion by constructing the orbitals according to the
symmetries of the system, the theory will become size-inconsistent, or in other words it will be dependent
on the particular representation employed. Thus one might arrive at a paradox, where in the 
self-interaction of $N$ hydrogen atoms arranged on a regular lattice of large lattice spacing (in such a way that 
there is no interaction between the atoms) vanishes, since the SIC of a Bloch state vanishes for $N\rightarrow\infty$.
However the SIC for an individual H atom, when calculated using atomic-like orbitals, 
accounts for essentially all the ground-state energy error \cite{PZSIC}. 
Therefore a size-consistent theory of SIC DFT must look for a scheme where a unitary transformation
of the occupied orbitals, which minimizes the SIC energy is performed. This idea is at the foundation of
all modern implementations of SIC. 

Significant progress towards the construction of a size-consistent SIC theory was made by Pederson, Heaton and
Lin, who introduced two sets of orbitals: localized orbitals $\phi_n^\sigma$ minimizing 
$E_\mathrm{xc}^\mathrm{SIC}$ and canonical (Kohn-Sham) de-localized orbitals $\psi_n^\sigma$ \cite{lin1,lin2,lin3}.
The localized orbitals are used for defining the densities entering into the effective potential (\ref{KSSIC-pot}),
while the canonical orbitals are used for extracting the Lagrangian multipliers $\epsilon_{nm}^{\sigma,\mathrm{SIC}}$, 
which are then associated to the KS eigenvalues. The two sets are related by unitary transformation 
$\psi_n^\sigma=\sum_m{\cal M}_{nm}^\sigma\phi_m^\sigma$, and one has two possible strategies for minimizing 
the total energy.

The first consists in a direct minimization with respect to the localized orbitals $\phi_n^\sigma$,
i.e. in solving equation (\ref{KSSIC-orth}) when we replace $\psi$ with $\phi$ and the orbital
densities entering the definition of the one-particle potential (\ref{KSSIC-pot}) are simply 
$\rho_n^\sigma=|\phi^\sigma_n|^2$. In addition the following minimization condition must be satisfied
\begin{equation}
\langle\phi_n^\sigma|v^\mathrm{SIC}_n-v^\mathrm{SIC}_m|\phi_m^\sigma\rangle=0\;,
\label{Mini}
\end{equation}
where $v^\mathrm{SIC}_n=u([\rho_n]; {\bf r})+v_\mathrm{xc}^{\sigma,\mathrm{LDA}}([\rho_n^\uparrow,0]; {\bf r})$. 
An expression for the gradient of the SIC functional, which also constrains the orbitals to be orthogonal
has been derived \cite{umigar} and applied to atoms and molecules with a mixture of successes and bad failures 
\cite{scuseria1,scuseria2,kummel1}.

The second strategy uses the canonical orbitals $\psi$ and seeks the minimization of the SIC energy
by varying both the orbitals $\psi$ and the unitary transformation ${\cal M}$. The corresponding set of equations is
\begin{equation}
H_n^\sigma\psi_n^\sigma=(H_0^\sigma+\Delta v^\mathrm{SIC}_n)\psi_n^\sigma=
\sum_m\epsilon_{nm}^{\sigma,\mathrm{SIC}}\psi_m^\sigma\;,
\label{KSSIC-pro}
\end{equation}
\begin{equation}
\psi^\sigma_n=\sum_m{\cal M}_{nm}\phi_m^\sigma\;,
\label{SIC-Rot}
\end{equation}
\begin{equation}
\Delta v^\mathrm{SIC}_n=\sum_m{\cal M}_{nm}v^\mathrm{SIC}_m\frac{\phi_m^\sigma}{\psi_n^\sigma}\;,
\label{DSIC}
\end{equation}
where $H_0^\sigma$ is the standard LDA Hamiltonian (without SIC). Thus the SIC potential for the
canonical orbitals appears as a weighted average of the SIC potential for the localized
orbitals. The solutions of the set of equations (\ref{KSSIC-pro}) is somehow more appealing than
that associated to the localized orbitals since the canonical orbitals can be constructed in a way
that preserves the system's symmetries (for instance translational invariance).

A convenient way for solving the equation (\ref{KSSIC-pro}) is that of using the so called ``unified 
Hamiltonian'' method \cite{lin1}. This is defined as (we drop the spin index $\sigma$)
\begin{equation}
H_\mathrm{u}=\sum_n^\mathrm{occup}\hat{P}_nH_0\hat{P}_n+
\sum_n^\mathrm{occup}(\hat{P}_nH_n\hat{Q}+\hat{Q}H_n\hat{P}_n)+
\hat{Q}H_0\hat{Q}\;,
\label{Unif}
\end{equation}
where $\hat{P}_n=|\psi_n^\sigma\rangle\langle\psi_n^\sigma|$ is the projector over the occupied orbital 
$\psi_n^\sigma$, and $\hat{Q}$ is the projector over the unoccupied ones $\hat{Q}=1-\sum_n^\mathrm{occup}\hat{P}_n$.
The crucial point is that the diagonal elements of the matrix $\epsilon_{nm}^{\sigma,\mathrm{SIC}}$ and
their corresponding orbitals $\psi_n^\sigma$ are respectively eigenvalues and eigenvectors
of $H_\mathrm{u}$, from which the whole $\epsilon_{nm}^{\sigma,\mathrm{SIC}}$ can be 
constructed. Finally, and perhaps most important,
at the minimum of the SIC functional, the canonical orbitals diagonalize the matrix 
$\epsilon_{nm}^{\sigma,\mathrm{SIC}}$, whose eigenvalues can now be interpreted as an analogue 
of the Kohn-Sham eigenvalues \cite{lin3}. 

It is also interesting to note that an alternative way for obtaining orbital energies is that of constructing 
an effective SI-free local potential using the Krieger-Li-Iafrate method \cite{KLI}. This has been recently 
explored by several groups \cite{garza,patch,martin}

When applied to extended systems the SIC method demands considerable additional computational 
overheads over standard LDA. Thus for a long time it has not encountered 
the favor of the general solid state community. In the case of solids the price to pay for not using canonical 
orbitals is enormous since the Bloch representation should be abandoned and in principle infinite 
cells should be considered. For this reason the second minimization scheme, in which the canonical 
orbitals are in a Bloch form, is more suitable. In this case for each ${\bf k}$-vector one can derive an 
equation identical to equation (\ref{KSSIC-pro}), where $\epsilon_{nm}^{\sigma,\mathrm{SIC}}=
\epsilon_{nm}^{\sigma,\mathrm{SIC}}({\bf k})$ and $n$ is simply the band index \cite{hhl}.
The associated localized orbitals $\phi$ for instance can be constructed as Wannier functions
and the minimization scheme proceeds in a similar way to that done for molecules. 

The problem here is that in practice, the cell needed to describe the localized states $\phi$ may be considerably
larger than the primitive unit cell. This is not the case for ionic insulators \cite{hhl}, where the localized orbitals
are well approximated by atomic orbitals. Such a simplification is however not valid in general. For example supercells as large
as 500 atoms have been considered for describing the localized $d$ shells in transition metals
oxides \cite{svane1,svane2,arai}. Despite these difficulties the SIC scheme has been applied
to a vast range of solid state systems with systematic improvement over LDA. These include,
in addition to transition metals monoxides \cite{svane1,arai,temm1}, rare-earth  materials
\cite{rareearth}, diluted magnetic semiconductors \cite{temm4},
Fe$_3$O$_4$ \cite{temm5}, heavy elements compounds \cite{heavy}, just to name a few.

In order to make the SIC method more scalable several approximations have been proposed. 
One possibility is that of incorporating part of the SIC into the definition of the pseudopotentials
\cite{pseudozunger}. The idea consists in subtracting the atomic SI from the free atom, 
and then transferring the resulting electronic structure
to the definition of a standard norm-conserving pseudopotential. 
This approximation is sustained by the fact that the transformation ${\cal M}$, which relates
canonical and localized orbitals does hardly mix core and valence states \cite{lin3}. Thus 
the SIC contribution to the total energy can be separated into the contributions from the core and 
the valence and in first approximation the latter can be neglected \cite{vogl}. 
The benefit of this method is that translational invariance is regained 
and the complicated procedure of minimizing ${\cal M}$ is replaced by a 
pseudopotential calculation. 

A further improvement over the pseudopotential method was recently presented by Vogel and co-workers 
\cite{voeg1,voeg2,voeg3} and then extended by Filippetti and Spaldin \cite{alessio}.
The method still assumes separability between the core and the valence contributions to the
SIC, but now the SIC for the valence electrons is approximated by an atomic-like 
contribution, instead of being neglected. This atomic SIC (ASIC) scheme is certainly a 
drastic approximation, since it implicitly assumes that the transformation ${\cal M}$ minimizing 
the SIC functional leads to atomic like orbitals. 

In the work of Vogel this additional contribution is not
evaluated self-consistently for the solid, while the implementation of Filippetti assumes a linear
dependance of the SIC over the orbital occupation. In spite of the approximations involved,
the method has been applied successfully to a range of solids including II-VI semiconductors and
nitrites \cite{voeg1,voeg2,alessio}, transition metal monoxides \cite{alessio,alessio2},
silver halides \cite{voeg3}, noble metal oxides \cite{alessio3}, 
ferroelectric materials \cite{alessio,alessio4,alessio5}, high-k materials \cite{alessio6}
and diluted magnetic semiconductors \cite{FSS1,FSS2}.
Interestingly most of the systems addressed by the ASIC method are characterized
by semi-core $d$ orbitals, for which an atomic correction looks appropriate, and a similar
argument is probably valid for ionic compounds as recently demonstrated for the case of 
SiC \cite{SiC}.

Here we further investigate the self-consistent ASIC method \cite{alessio} by 
examining both finite and extended systems, and by critically considering whether a scaling
factor, additional to the orbital occupation, is needed for reproducing the correct single 
particle spectrum. 


\section{Formalism and Implementation}
In this section we derive the fundamental equations of the ASIC method, while looking closely at
the main approximations involved in comparison to the fully self-consistent SIC approach. Our practical implementation
is also described.


\subsection{The ASIC potential}
The starting point of our analysis is the SIC Schr\"odinger-like equation (\ref{KSSIC-pro}) for the canonical 
orbitals. Let us assume, as from reference \cite{vogl}, that the rotation ${\cal M}$ transforming
localized orbitals (to be determined) into canonical orbitals (see equation (\ref{SIC-Rot})) does not mix core and
valence states. We also assume that core electrons are well localized into atomic-like
wave-functions and that they can be effectively described by a norm-conserving pseudopotential.

Let us now assume that ${\cal M}$ is known and so are the localized orbitals $\phi_m^\sigma$. In this
case the canonical orbitals diagonalize $\epsilon_{nm}^{\sigma,\mathrm{SIC}}$ and the 
equation (\ref{KSSIC-pro}) simply reduces to 
\begin{equation}
(H_0^\sigma+\Delta v^\mathrm{SIC}_n)\psi_n^\sigma=
\epsilon_{n}^{\sigma,\mathrm{SIC}}\psi_n^\sigma\;, 
\label{KSSIC-diag}
\end{equation}
with $\Delta v^\mathrm{SIC}_n$ defined in equation (\ref{DSIC}). The Hamiltonian 
$H_0^\sigma+\Delta v^\mathrm{SIC}_n$ can be then re-written in a convenient form
as
\begin{equation}
H_0^\sigma+\Delta v^\mathrm{SIC}_n=H_0^\sigma+\sum_m v^\mathrm{SIC}_m \hat{P}_m^\phi\;,
\label{HSIC-conv}
\end{equation}
where $v^\mathrm{SIC}_m$ is the self-interaction potential for the localized orbital $\phi_m^\sigma$,
and $\hat{P}_m^\phi$ is the projector over the same state
\begin{equation}
\hat{P}_m^\phi\psi_n^\sigma({\bf r})=\phi_m^\sigma({\bf r})\:\int\mathrm{d}^3{\bf r}^\prime\:
\psi_n^\sigma({\bf r}^\prime)\phi_m^{\sigma\dagger}({\bf r}^\prime)=
\phi_m^\sigma({\bf r})\langle\phi_m^\sigma|\psi_n^\sigma\rangle\;.
\label{PhiProj}
\end{equation}

Two main approximations are then taken in the ASIC approach \cite{voeg1,alessio}. First the
localized states $\phi_m^\sigma$ are assumed to be atomic-like orbitals $\Phi_m^\sigma$ (ASIC orbitals) 
and the SIC potential is approximated as
\begin{equation}
\sum_m v^\mathrm{SIC}_m({\bf r}) \hat{P}_m^\phi\:\rightarrow\:
\alpha\:\sum_m \tilde{v}^{\sigma\mathrm{SIC}}_m({\bf r}) \hat{P}_m^\Phi
\;, 
\label{PSIC-appr}
\end{equation}
with $\tilde{v}^{\sigma\mathrm{SIC}}_m({\bf r})=u([\rho_m]; {\bf r})+v_\mathrm{xc}^{\sigma,\mathrm{LDA}}([\rho_m^\uparrow,0]; {\bf r})$ and
$\rho_m^\sigma=|\Phi_m^\sigma|^2$, $\hat{P}_m^\Phi$ is the projector of equation (\ref{PhiProj}) obtained by
replacing the $\phi$'s with the ASIC orbitals $\Phi$, and $\alpha$ is a scaling factor. Note that the orbitals $\Phi_m$ are not
explicitly spin-dependent and one simply has $\Phi_m^\sigma$=$\Phi_m\:p_m^\sigma$ with $p_m^\sigma$ the
orbital occupation ($p_m^\sigma=0,1$). The factor $\alpha$ is an empirical factor, which accounts for the
particular choice of ASIC orbitals. This first approximation is expected to be accurate for systems retaining
an atomic-like charge density as in the case of small molecules. It is also formally exact in the one-electron
limit (for $\alpha=1$). In the case of extended solids the situation is less transparent, since in general the 
functions minimizing $E_\mathrm{xc}^\mathrm{SIC}$ are Wannier-like functions \cite{nicola}.

The second approximation taken in the ASIC method is that of replacing the non-local projector $\hat{P}_m^\Phi$ 
with its expectation value. The idea is that the SIC potential for the canonical orbitals $\Delta v^\mathrm{SIC}_n$
is formally a weighted average of the SIC potential for the localized orbitals $v_m^\mathrm{SIC}$. For the exact SIC
method the weighting factor is the non-local projector ${\cal M}_{nm}\frac{\phi_m^\sigma}{\psi_n^\sigma}$. This 
means that the SIC potential for a given canonical orbital $\psi_n$ is maximized in those regions where the overlap with
some of the localized orbitals $\phi_m$ is maximum. In the ASIC method such non-local projector is replaced
more conveniently by a scalar. In the original proposal by Voegl et 
al. \cite{voeg1,voeg2,voeg3} this was simply set to one. Here we consider the orbital occupation $p_m^\sigma$
of the given ASIC orbital $\Phi_m$, i.e. we replace $\hat{P}_m^\Phi$ with its expectation value
\begin{equation}
\hat{P}_m^\Phi\rightarrow\langle\hat{P}_m^\Phi\rangle=p_m^\sigma=\sum_nf_n^\sigma\langle\psi_n^\sigma|
\hat{P}_m^\Phi|\psi_n^\sigma\rangle\:,
\label{EVProj}
\end{equation}
where $f_n^\sigma$ is the occupation number of the Kohn-Sham orbital $\psi_n^\sigma$. The final form of
ASIC potential is then 
\begin{equation}
v_\mathrm{ASIC}^\sigma({\bf r})=\alpha\:\sum_m \tilde{v}^{\sigma\mathrm{SIC}}_m({\bf r}) p_m^\sigma\;.
\label{PSIC-appr-final}
\end{equation}

Let us now comment on the empirical scaling factor $\alpha$. In reference \cite{alessio} $\alpha$ was set to $1/2$ 
in order to capture eigenvalue relaxation. This choice however violates the one-electron limit of the SIC potential,
which is correctly reproduced for $\alpha=1$. We can then interpret $\alpha$
as a measure of the deviation of the ASIC potential from the exact SIC potential. Ultimately 
$\alpha$ reflects the deviation of the actual ASIC projectors $|\Phi\rangle\langle\Phi|$ from the
localized orbitals defining the SI corrected ground state. One then expects $\alpha$ to be
close to 1 for systems with an atomic-like charge density, and to vanish for metals, whose valence charge 
density resembles that of a uniform electron gas \cite{Norman}. 
Intermediate values are then expected for situations different from these two extremes, and we will show
that a values around $1/2$ describe well a vast class of mid- to wide-gap insulators.


\subsection{Implementation}

The final form of the SIC potential to subtract from the LSDA (local spin density approximation) 
one (equation (\ref{PSIC-appr-final})) is that 
of a linear combination of non-local pseudopotential-like terms. These are uniquely defined by the 
choice of exchange and correlation potential used and by ASIC orbitals $\Phi_m$. 
The practical way of constructing such potentials, i.e. the way of importing the
atomic SIC to the solid state, depends on the specific numerical implementation used for the DFT algorithm. 
At present only plane-wave implementations are available \cite{alessio, voeg1,voeg2,voeg3}, while here we 
present our new scheme based on the pseudo atomic orbital (PAO) \cite{MZeta} code {\it Siesta} \cite{siesta}. 

We start by solving the atomic all-electron SIC-LSDA equation for all the species involved in the solid state
calculation. Here we apply the original Perdew-Zunger (PZ-SIC) formalism \cite{PZSIC} and we neglect the small
non-orthogonality between the Kohn-Sham orbitals. Thus we obtain a set of SI corrected atomic orbitals $\Phi_m$, 
which exactly solve the atomic SIC-LSDA problem. The atomic orbitals $\Phi_m$ describing the valence electrons 
are then used to define the ASIC potentials $\tilde{v}_m^{\sigma\mathrm{SIC}}$
\begin{equation}
\tilde{v}^{\sigma\mathrm{SIC}}_m({\bf r})=u([\rho_m]; {\bf r})+v_\mathrm{xc}^{\sigma,\mathrm{LDA}}([\rho_m^\uparrow,0]; {\bf r})
\end{equation}
with $\rho_m^\sigma=|\Phi_m|^2$. 

At the same time a standard LSDA calculation for the same atoms is used to construct the pseudopotentials 
describing the core electrons. These are standard norm-conserving scalar relativistic Troullier-Martins
pseudopotentials \cite{MT} with nonlinear core corrections \cite{corr}. Thus we usually neglect the SIC over 
the core states, when constructing the pseudopotentials. This is justified by the fact that the eigenvalues for the 
SIC-LSDA-pseudoatom, i.e. for the free atom where the effects of core electrons are replaced by LSDA 
pseudopotentials but SIC is applied to the valence electrons, are in excellent agreement with 
those obtained by all-electron SIC-LSDA calculations \cite{voeg1}. 

The final step is that of recasting the ASIC potentials $\tilde{v}^\mathrm{SIC}_m({\bf r})$, which have a $-2/r$ asymptotic
behaviour, in a suitable non-local form. This is obtained with the standard Kleinman-Bylander \cite{KB} scheme
and the final ASIC potential (equation (\ref{PSIC-appr-final})) is written as
\begin{equation}
v_\mathrm{ASIC}^\sigma = \sum_{m} \frac{| \gamma_{m}^{\sigma} \rangle 
\langle \gamma_{m}^{\sigma} |}{C_{m}^{\sigma}}\;, 
\end{equation}
where the ASIC projectors are given by
\begin{equation}
\gamma_{m}^{\sigma}({\bf r}) = \alpha\:p^\sigma_m\:\tilde{v}_m^{\sigma\mathrm{SIC}}({\bf r})\Phi^\prime_{m}({\bf r})\:. 
\end{equation}
and the normalization factors are
\begin{equation}
C_{m}^{\sigma} =  \alpha\:p^\sigma_m\langle \Phi^\prime_{m}|\tilde{v}_m^{\sigma\mathrm{SIC}}|\Phi^\prime_{m}  \rangle\:.
\end{equation}

The orbital functions $\Phi_m^\prime$ are atomic-like functions with a finite range, which ensure that the ASIC
projectors $\gamma_m$ do not extend beyond that range. These are constructed in the same way as the {\it Siesta}
basis set orbitals, i.e. as solutions of the pseudo-atomic problem with an additional confining potential at the cutoff radius 
$r_\mathrm{cutoff}$ \cite{MZeta}. The choice of the appropriate cutoff radius for the SIC projectors should take into account the
two following requirements. On the one hand it should be sufficiently large to capture most of the
SIC corrections. A good criterion \cite{alessio} is that the SIC-LSDA contribution to the orbital energy of the free atom 
\begin{eqnarray}
\mathbf{\delta\varepsilon}^\sigma_{\mathrm{SIC}\:m} = \langle \Phi_m^\prime|
\tilde{v}^{\sigma\mathrm{SIC}}_m|\Phi_m^\prime\rangle
\end{eqnarray}
is reproduced within some tolerance. On the other hand the cutoff should be reasonably short so as not to change
the connectivity of the matrix elements of the PAO Hamiltonian. In other words we need to ensure that orbitals 
otherwise considered as disconnected in evaluating the various parts of the Hamiltonian matrix are not considered 
connected for the $v_\mathrm{ASIC}^\sigma$ matrix elements alone.

As a practical rule we set the cutoff radius for a particular orbital of a given atom to be either equal to the longest among
the cutoff radii of the PAO basis set for that particular atom (typically the first $\zeta$ of the lowest angular momentum),
or, if shorter, the radius at which $\mathbf{\delta\varepsilon}^\sigma_{\mathrm{SIC}\:m}<$~0.1mRy.
Typically, when reasonable cutoff radii (6 to 9 Bohr) are used,  we find that the atomic SIC-LSDA eigenvalues are
reproduced to within 1 to 5 mRy for the most extended shells and to within 0.1 mRy for more confined shells. 
Thus $\mathbf{\delta\varepsilon}^\sigma_{\mathrm{SIC}\:m}$ are rather well converged already for cutoff radii 
defined by a PAO energies shifts \cite{siesta} of around 20mRy, although usually smaller PAO energy shifts are necessary 
for highly converged total energy calculations.

Finally the matrix elements of the SIC potential are calculated as usual over the {\it Siesta} basis set. Additional
basis functions $\chi_m$ are constructed from the confined localized atomic orbitals described before using
the split-norm scheme \cite{MZeta,MZeta1,MZeta2}. The density matrix $\rho^\sigma$ is represented over such basis
$\rho_{\mu\nu}^\sigma$ and the orbital populations are calculated as
\begin{equation}
p_{m}^{\sigma} = \sum_{\mu\nu} S_{m \mu } \rho^\sigma_{\mu\nu}S_{\nu m}\;,
\end{equation}
where $S_{m \mu }$ is a matrix element of the overlap matrix. Note that in principle the orbital population
should be constructed for the atomic SIC orbital $\Phi_m$. However, we notice that $p_m^\sigma$ is rather insensitive 
to the specific choice of orbital, once this has a reasonable radial range. For practical numerical reasons in the present 
implementation, we always use the orbital populations projected onto the basis set sub-space consisting of the most extended first-$\zeta$ orbitals of the atomic species involved.
The matrix elements of the SIC potential are simply
\begin{equation}
(v_\mathrm{ASIC}^\sigma)_{\mu\nu}=
\sum_{m} \frac{\langle\chi_\mu| \gamma_{m}^{\sigma} \rangle 
\langle \gamma_{m}^{\sigma} |\chi_\nu\rangle}{C_{m}^{\sigma}}\:,
\end{equation}
and the ASIC-KS equation takes the final form
\begin{equation}
\left[-\frac{1}{2}\boldsymbol{\nabla}^2+v_\mathrm{PP}+u+v_\mathrm{xc}^{\sigma,\mathrm{LSDA}}-
v^\sigma_\mathrm{ASIC}\right]\psi_n^\sigma=\epsilon_{n}^{\sigma,\mathrm{SIC}}\psi_n^\sigma\;.
\label{KS-ASIC-final}
\end{equation}
with $v_\mathrm{PP}$ the pseudopotential.


\subsection{Total Energy}

The energy corresponding to the SIC-LSDA functional is given by \cite{PZSIC}
\begin{equation}
E^\mathrm{SIC}[\rho^\uparrow,\rho^\downarrow]=E^\mathrm{LSDA}[\rho^\uparrow,\rho^\downarrow]-
\sum_{n\sigma}^\mathrm{occ.}(U[\rho_n^\sigma]+E_\mathrm{xc}^\mathrm{LSDA}[\rho_n^\sigma,0])\;,
\label{TotESIC}
\end{equation}
where 
\begin{equation}
U[\rho_n^\sigma]=\int\mathrm{d}^3{\bf r}\:\frac{1}{2}\rho_n({\bf r})u([\rho_n]; {\bf r})\;,
\label{Hartree}
\end{equation}
\begin{equation}
E_\mathrm{xc}^\mathrm{LSDA}[\rho_n^\sigma,0])=\int\mathrm{d}^3{\bf r}\:\rho_n({\bf r})\mathcal{E}^\mathrm{LSDA}_\mathrm{xc}
([\rho_n]; {\bf r})\;,
\label{XC}
\end{equation}
with $\mathcal{E}^\mathrm{LSDA}_\mathrm{xc}$ the LSDA exchange and correlation energy density.
The orbital densities entering in the SI term are those associated to the local orbitals
$\phi$. As we have already mentioned, this functional needs to be minimized with respect to the $\phi$'s,
which are an implicit function of the total spin density $\rho^\sigma$. In the ASIC approximation these orbitals
are not minimized, but taken as atomic functions. This means that in the present form the
theory is not variational, in the sense that there is no functional related to the KS equation (\ref{KS-ASIC-final})
by a variational principle. With this in mind we adopt the expression of equation (\ref{TotESIC}) as a
suitable energy, where the orbital densities are those given by the ASIC orbitals
\begin{equation}
\rho_m^\sigma({\bf r})=p_m^\sigma\:|\Phi_m|^2\;.
\label{OrbDen}
\end{equation}

In our implementation the LSDA KS energy $E^\mathrm{LSDA}$ is directly available as calculated in the {\it Siesta} 
code \cite{siesta} and thus, only the second term of equation (\ref{TotESIC}) needs to be calculated. This is easily 
done by calculating both $U$ and $E_\mathrm{xc}^\mathrm{LSDA}$ on an atomic radial grid for each atomic orbital in the system.


\subsection{ASIC and LDA+$U$}

We now compare our ASIC method with another atomic-like correction to LSDA, namely the LDA+$U$ method
\cite{AnisimovZaanen,AnisimovLongPaper}. In LDA+$U$ one replaces  the LSDA 
exchange and correlation energy associated to the ``correlated'' orbitals ($d$ or $f$ shells), with the Hubbard-$U$ energy. 
Thus the functional becomes
\begin{equation}
E^\mathrm{LDA+U}[\rho({\bf r})]=E^\mathrm{LSDA}[\rho({\bf r})]+E^\mathrm{U}[\{p_m^\sigma\}]-E^\mathrm{DC}[\{p_m^\sigma\}]\;,
\label{LDAU}
\end{equation}
where the Hubbard energy $E^\mathrm{U}$ and the double counting term $E^\mathrm{DC}$ depend on the orbital 
populations $p_m^\sigma$ of the correlated orbitals. Several forms for the LDA+$U$ functional have been proposed
to date. A particularly simple and transparent one \cite{degir,Dude}, which is also rotationally invariant, redefines the $U$ 
parameter as an effective parameter $U_\mathrm{eff}=U-J$ and the functional takes the form
\begin{equation}
E^\mathrm{U}-E^\mathrm{DC}=\frac{U_\mathrm{eff}}{2}\sum_I\sum_{m\:\sigma}\left[p_{mm}^{I\:\sigma}-\sum_n\:p_{mn}^{I\:\sigma}\:
p_{nm}^{I\:\sigma}\right]
\label{LDAURot}
\end{equation}
where in we separate out the index for the atomic position $I$ from the magnetic quantum number $m$, and 
introduce the off-diagonal populations $p_{mn}^{I\:\sigma}=\sum_\alpha f_\alpha^\sigma\langle\psi_\alpha^\sigma|
\hat{P}_{mn}^{I\:\Phi}|\psi_\alpha^\sigma\rangle$ with $\hat{P}_{mn}^{I\:\Phi}=|\Phi_m^I\rangle\langle\Phi_n^I|$.
Note that the LDA+$U$ functional is SI free for those orbitals that are corrected. 

Although a rotationally invariant form of the ASIC potential can be easily derived, we assume here for simplicity
that the system under consideration is rotationally invariant, or alternatively that we have carried out a rotation, which
diagonalizes the $p_{mn}^{I\:\sigma}$ matrix. In this case the energy becomes simply
\begin{equation}
E^\mathrm{U}-E^\mathrm{DC}=\frac{U_\mathrm{eff}}{2}\sum_{Im\:\sigma}\:p_{m}^{I\:\sigma}\left[1-\:p_{m}^{I\:\sigma}
\right]\;,
\label{LDAURotation}
\end{equation}
with $p_{m}^{I\:\sigma}=p_{mm}^{I\:\sigma}$. It is then easy to compute the KS potential
\begin{equation}
v^\mathrm{LDA+U}=v^\mathrm{LSDA}+\sum_{Im\:\sigma}U_\mathrm{eff}\left[\frac{1}{2}-\:p_{m}^{I\:\sigma}\right]
\hat{P}^{I\Phi}_m\;,
\label{LDAUPot}
\end{equation}
and the orbital energy
\begin{equation}
\epsilon^{I\sigma}_m=\frac{\partial E}{\partial p^{I\sigma}_m}=\epsilon^{I\sigma\:\mathrm{LSDA}}_m
+U_\mathrm{eff}\left(\frac{1}{2}-p_m^{I\sigma}\right)\;.
\label{LDAUEne}
\end{equation}

These need to be compared with the ASIC potential (equation (\ref{PSIC-appr-final})) and orbital energy 
\begin{equation}
\epsilon^{I\sigma}_m=\epsilon^{I\sigma\:\mathrm{LSDA}}_m
-\alpha\:p^{I\sigma}_m\langle \Phi^{I\prime}_{m}|\tilde{v}_{Im}^{\sigma\mathrm{SIC}}|\Phi^{I\prime}_{m}\rangle\;,
\label{SICEne}
\end{equation}
where the last term follows from $\frac{\partial E}{\partial p^{I\sigma}_m}=C^{I\sigma}_m$ and from equation (\ref{TotESIC}). 
The main difference between the ASIC and LDA+$U$ method is in the way in which 
unoccupied states are handled. In fact, while LDA+$U$ corrects unoccupied states and pushes the orbital energies
upwards by $\sim U_\mathrm{eff}/2$, ASIC operates only on occupied states, that are shifted towards lower energies by
$C^{I^\sigma}_m$. This reflects the fact that the SIC is defined only for occupied KS orbitals.
An important consequence is that the opening of bandgaps in the electronic structures, 
one of the main features of both the LDA+$U$ and ASIC schemes, is then driven by two different 
mechanisms. On the one hand in LDA+$U$, gaps open up since occupied and unoccupied states are corrected in opposite 
directions leading to a gap of $\sim U_\mathrm{eff}$. On the other hand ASIC is active only over occupied states and gaps open
only if occupied and unoccupied bands have large differences in their projected atomic orbital content. Thus one should not
expect any corrections for covalent materials where conduction and valence bands are simply bonding and
antibonding states formed by the same atomic orbitals. This is for instance the case of Si and Ge. In contrast ASIC will
be extremely effective for more ionic situations, where the orbital contents of conduction and valence bands are
different.

Finally, by comparing the corrections to the orbital energy of a fully occupied state, one finds
\begin{equation}
U=2\alpha\;\langle \Phi^{I\prime}_{m}|\tilde{v}_{Im}^{\sigma\mathrm{SIC}}|\Phi^{I\prime}_{m}\rangle\:,
\label{UvsASIC}
\end{equation}
which establishes an empirical relation between the Hubbard energy and the ASIC correction. Since $U$
is sensitive to the chemical environment due to screening \cite{degir}, while all the other quantities are 
uniquely defined by an atomic calculation, we can re-interpret the parameter $\alpha$ as empirically 
describing the screening from the chemical environment within the ASIC scheme.


\section{Results: Extended systems}

The test calculations that we present in this work are for two classes of materials: extended and finite. 
First we investigate how our implementation performs in the solid state. In particular we discuss the 
r\^ole of the parameter $\alpha$ in determining the bandstructure of several semiconductors, considering both
the KS band-gap and the position of bands associated with tightly bound electrons.


\subsection{Estimate of $\alpha$ for semiconductors}

The quasi-particle band gap $E_g$ in a semiconductor is defined as the difference between its ionization potential $I$ and 
electron affinity $A$. These can be rigorously calculated from DFT as the HOMO energy respectively of the neutral and negatively
charged systems. This actual gap cannot be directly compared with the KS band-gap $E_g^\mathrm{KS}$, defined as the
difference between the orbital energy of the HOMO and LUMO states of the neutral system. In fact, the presence of a derivative 
discontinuity in the DFT energy as a function of the electron occupation establishes the following rigorous relation \cite{PPLB,ShamSlu}
\begin{equation}
E_g=E^\mathrm{KS}_g+\Delta_\mathrm{xc}\:,
\label{realgap}
\end{equation}
with
\begin{equation}
\Delta_\mathrm{xc}=\lim_{\omega\rightarrow0^+}\left\{\frac{\delta E_\mathrm{xc}[n]}{\delta n}\left|_{N+\omega}-\frac{\delta E_\mathrm{xc}[n]}
{\delta n}\right|_{N-\omega}\right\}_{n_N}\:.
\label{DerDisc}
\end{equation}
This is valid even for the exact XC potential, and therefore in principle one has to give up KS
bandstructures as a tool for evaluating semiconductor band-gaps. The size of $\Delta_\mathrm{xc}$ is however not known for
real extended systems and the question of whether most of the error in determining  $E_g$ from $E_g^\mathrm{KS}$ is
due to the approximation in the XC potential or due to the intrinsic $\Delta_\mathrm{xc}$ is a matter of debate. 

In general SI-free potentials bind more than LSDA and one expects larger gaps. Surprisingly, functionals based on 
exact exchange, provide KS gaps rather close to the experimental values \cite{gor,gross1}. The reason 
for such a good agreement is not fully understood, but it is believed that the exact KS gaps should be smaller than the 
actual ones. 

With this in mind, we adopt a heuristic approach and we use the KS band-gap as a quality indicator for
interpreting the parameter $\alpha$ and for providing its numerical value for different classes of solids. Here we 
investigate the dependence of $E^\mathrm{KS}_g$ over $\alpha$ and we determine the value for $\alpha$ yielding 
the experimental band-gap. Assuming that $\Delta_\mathrm{xc}$ does not vary considerably across the materials 
investigated, this will allow us to relate $\alpha$ to the degree of localization in a semiconductor and to extract the value
useful for ASIC to be an accurate single-particle theory.

\begin{figure}[ht]
\includegraphics[width=0.43\textwidth]{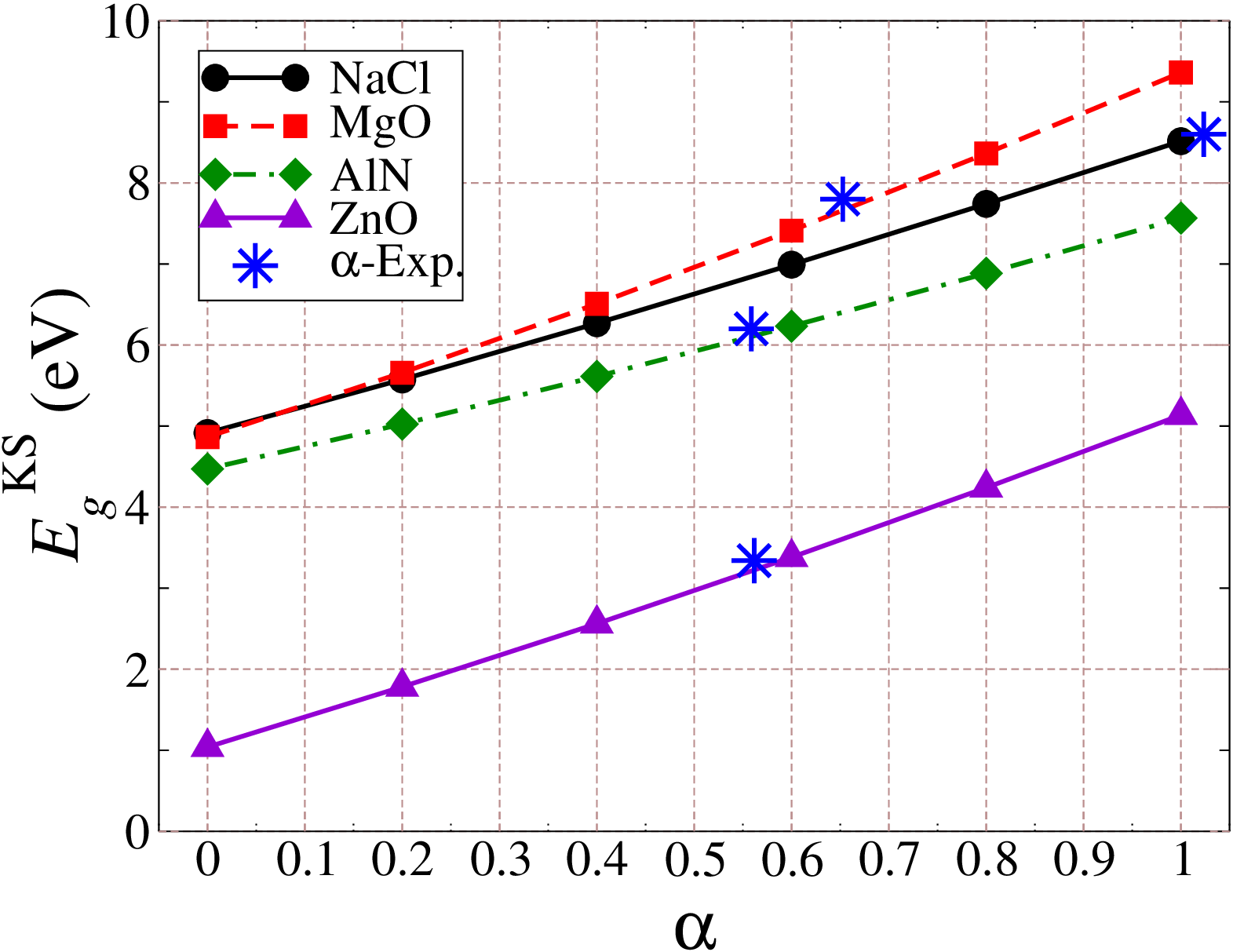}
\caption{\label{Fig1}Calculated band-gap for NaCl, MgO,AlN and ZnO as a function of the parameter $\alpha$. 
$\alpha=0$ is the LSDA value and $\alpha=1$ accounts for complete atomic SIC. The lattice parameters
used for the calculations are either the equilibrium LSDA or the experimental when available.}
\end{figure}

In figure \ref{Fig1} we present the band-gap of four representative semiconductors as a function of $\alpha$ together with
the value needed to reproduce the experimental band-gap. LSDA corresponds to $\alpha=0$ and while $\alpha=1$
accounts for the full ASIC. 
In general $E_g^\mathrm{KS}$ increases as $\alpha$ increases, as a result of the stronger SIC. The
$E_g^\mathrm{KS}(\alpha)$ curve is almost linear with a slope, which appears to be material-specific.

For the most ionic compound, NaCl, the experimental gap is reproduced almost exactly by $\alpha=1$, i.e. by the
full ASIC. This is somehow expected since the charge density of solid NaCl is rather close
to a superposition of the Na$^+$ and Cl$^-$ ionic charge densities. In this case of strongly localized charge densities
the ASIC approximation is rather accurate yielding results substantially identical to those obtained with full 
self-consistent PZ-SIC \cite{hhl}. Indeed earlier calculations for LiCl \cite{hhl}  demonstrate that the SIC band-structure 
is rather insensitive of the localized orbitals $\phi$ once these have an atomic-like form. 

For the other compounds the localized orbitals $\phi$'s are not necessarily atomic-like functions and deviations
from $\alpha=1$ are expected. Interestingly we find that, for all the materials investigated, a value of around 1/2
reproduces the experimental band-gap rather accurately. As an illustration, in table \ref{TabGap} we compare the
experimental band-gap $E_g^\mathrm{exp}$ to the calculated $E_g^\mathrm{KS}$ for ASIC ($\alpha=1$) and LDA,
for several semiconductors ranging from ionic salts to wide-gap II-VI and III-V semiconductors. 
We also report the value of $\alpha=\alpha^*$ needed for $E_g^\mathrm{exp}=E_g^\mathrm{KS}$.
\begin{table}[h]
\caption{\label{TabGap}Experimental $E_g^\mathrm{exp}$ and KS $E_g^\mathrm{KS}$ band-gap (in eV) for a number of
semiconductors. $E_g^\mathrm{KS}$ are calculated with both LSDA and ASIC ($\alpha=1$). In the last
column we report the value of $\alpha=\alpha^*$ needed for $E_g^\mathrm{exp}=E_g^\mathrm{KS}$. The lattice parameters
used for the calculations are either the equilibrium LSDA or the experimental when available (in \AA). RS=rocksalt, WZ=wurtzite,
ZB=zincblende. The value for the experimental gaps are from the literature: 
$a$ \cite{LiClGap}, 
$b$ \cite{NaClGap}, 
$c$ \cite{KClGap}, 
$d$ \cite{MgOGap}, 
$e$ \cite{SrOGap}, 
$f$ \cite{AlNGap}, 
$g$ \cite{GaNGap}, 
$h$ \cite{InNGap}, 
$i$ \cite{ZnOGap}}
\begin{ruledtabular}
\begin{tabular}{lccccc}
Solid&Structure&$E_g^\mathrm{exp}$&$E_g^\mathrm{KS-LSDA}$&$E_g^\mathrm{KS-ASIC}$&$\alpha$\\
\hline
LiCl   & RS ($a=$5.13) & 9.4$^a$ & 6.23 & 9.76 & 0.89\\
NaCl & RS ($a=$5.63) & 8.6$^b$ & 4.91 & 8.51 & 1.02\\
KCl    & RS ($a=$6.24) & 8.5$^c$ & 4.90 & 8.51 & 0.99\\
MgO   & RS ($a=$4.19) & 7.8$^d$ & 4.86 & 9.36 & 0.65\\
CaO   & RS ($a=$4.74) & 7.08$^d$ & 4.93 & 9.28 & 0.49\\
SrO   & RS ($a=$5.03) & 5.89$^e$ & 4.20 & 7.80 & 0.47\\
AlN   & WZ ($a$=3.11, $c$=4.98) & 6.20$^f$ & 4.47 & 7.56 & 0.56\\
GaN   & WZ ($a$=3.16, $c$=5.13) & 3.39$^g$ & 2.21 & 5.03 & 0.44\\
InN   & WZ ($a$=3.54, $c$=5.70) & 0.7$^h$ & 0.09 & 2.09 & 0.45\\
ZnO   & WZ ($a$=3.23, $c$=5.19) & 3.43$^i$ & 0.85 & 5.13 & 0.57\\
ZnS   & ZB ($a=$5.40) & 3.78$^i$ & 2.47 & 4.90 & 0.53\\
ZnSe   & ZB ($a=$5.63) & 2.82$^i$ & 1.77 & 3.53 & 0.58\\
\end{tabular}
\end{ruledtabular}
\end{table}

Clearly for all the strongly ionic compounds (LiCl, NaCl and KCl) the full ASIC correction $\alpha=1$ reproduces
quite accurately the experimental gap and agrees with previous self-consistent SIC calculations \cite{NaClSIC}. 
For all the other compounds a value of around 1/2 is always adequate, confirming the initial choice of 
Filippetti and Spaldin. For these materilas we do not find any particular regularity. In general $\alpha$ is large
when the experimental gap is large, however there is no direct connection between $\alpha$ and the ionicity
or covalency of a compound. In fact, the improvement of the band-gap is not simply due to a rigid shift of
the valence band, but usually corresponds to a general improvement of the whole quasi-particle spectrum.
Examples for ZnO and GaN will be presented in the next section. 

As a further proof of this point in table \ref{TabWid} we present the valence band-width for the semiconductors
investigated as calculated from LSDA $\Delta E_v^\mathrm{LSDA}$ and ASIC for both $\alpha=1$ 
($\Delta E_v^\mathrm{ASIC_1}$) and $\alpha=\alpha^*$ ($\Delta E_v^\mathrm{ASIC_{\alpha^*}}$). We also report
the experimental values $\Delta E_v^\mathrm{exp}$ whenever available, although a direct comparison with experiments 
is difficult, since these values are rather imprecise and sometimes not known. The general feature is that ASIC produces 
only minor corrections over LSDA, and that these corrections do not follow a generic trend. Thus, while for the nitrites
ASIC always increases the band-width, it does just the opposite for KCl, SrO and CaO.
\begin{table}[h]
\caption{\label{TabWid}Valence band experimental bandwidth $\Delta E_v^\mathrm{exp}$ compared with those
obtained from ASIC ($\alpha=1$) $\Delta E_v^\mathrm{ASIC_1}$, LSDA ($\Delta E_v^\mathrm{LSDA}$) and
ASIC with the optimal $\alpha=\alpha^*$ from table \ref{TabGap} $\Delta E_v^\mathrm{ASIC_{\alpha^*}}$ for a number of
semiconductors. The lattice parameters used for the calculations are either the equilibrium LSDA or the 
experimental when available (in \AA). The experimental values are from the literature (last column).}
\begin{ruledtabular}
\begin{tabular}{lcccccr}
Solid&$\Delta E_v^\mathrm{exp}$&$\Delta E_v^\mathrm{ASIC_1}$&$\Delta E_v^\mathrm{LSDA}$&$\Delta E_v^\mathrm{ASIC_{\alpha^*}}$&Reference\\
\hline
LiCl     & 4-5        & 3.52 & 3.06 &3.51 & \cite{NaClSIC}  \\
NaCl   & 1.7-4.5 & 2.11 & 2.06 & 2.11 & \cite{NaClSIC} \\
KCl      & 2.3-4.3 & 1.09 & 1.21 & 1.09 & \cite{NaClSIC}  \\
MgO    & 3.3-6.7        & 5.16 & 4.83 & 5.06 & \cite{MgOWD,CaOWD} \\
CaO    &   0.9            & 2.72 & 2.89 & 2.82 & \cite{CaOWD} \\
SrO     &               & 2.21 & 2.53 & 2.39 & \\
AlN     & 6.0        & 7.44 & 6.27 & 6.92 & \cite{AlNWD}  \\
GaN   & 7.4        & 8.42 & 7.33 & 7.85  & \cite{GaNWD}  \\
InN     &   6.0            & 6.66 & 6.01 & 6.34 & \cite{InNWD} \\
ZnO    & $\sim$5 & 5.66 & 4.77 & 5.54 & \cite{ZnOWD} \\
ZnS    &    5.5          & 6.49 & 5.57 & 6.05 & \cite{ZnSWD} \\
ZnSe  &    5.6          & 7.14 & 5.35 & 6.38 & \cite{ZnSWD}\\
\end{tabular}
\end{ruledtabular}
\end{table}


\subsection{Wide-gap semiconductors: ZnO and GaN}

\begin{figure}[ht]
\includegraphics[width=0.45\textwidth]{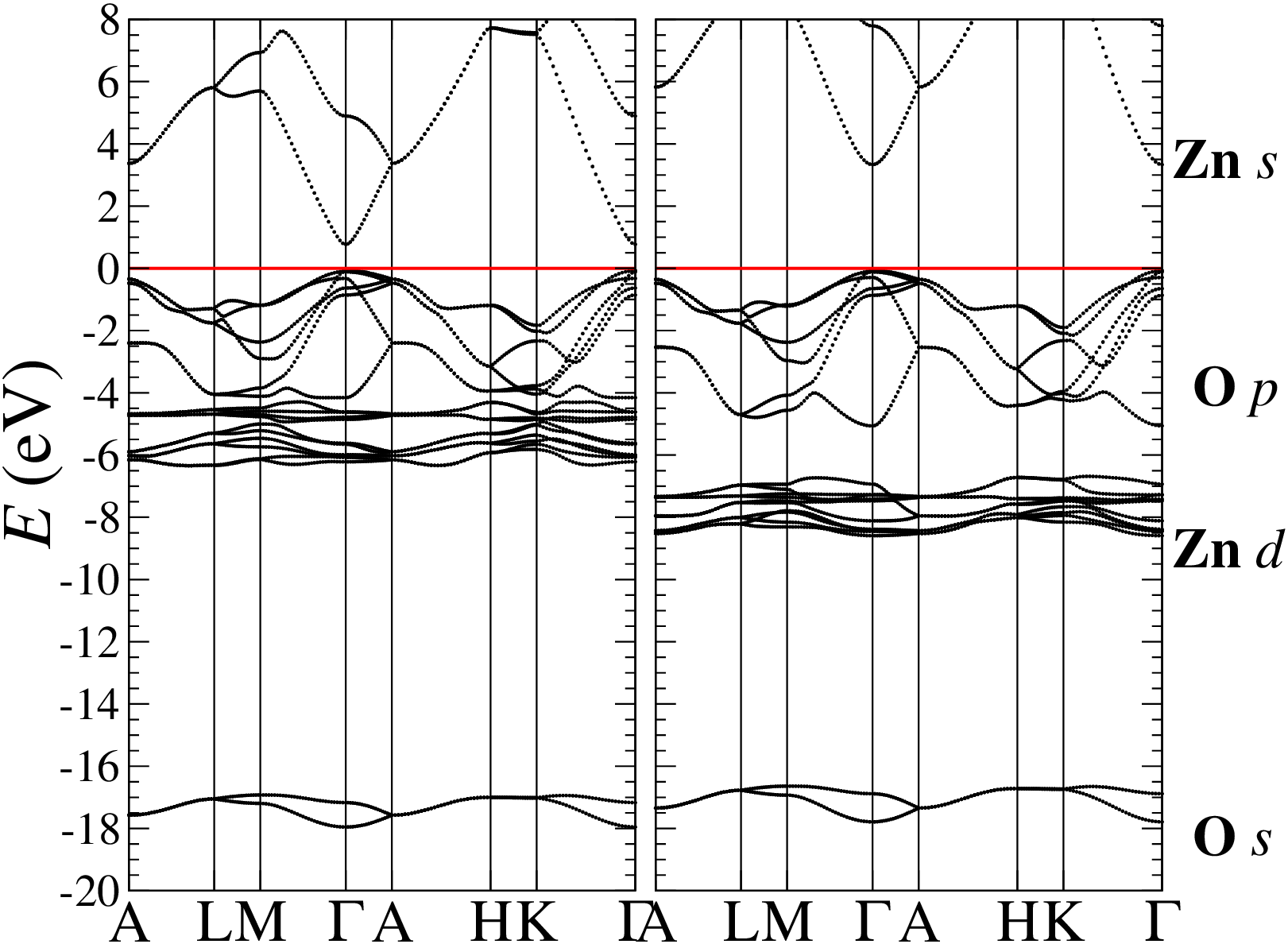}
\caption{\label{ZnO.bands}Calculated band structure of wurtzite ZnO obtained from LSDA and ASIC. Owing to the 
ionic character of ZnO each group of bands can be clearly labeled according to a single, dominant orbital 
character as shown. The VBT is aligned at 0~eV.}
\end{figure}

Having established $\alpha=1/2$ as an appropriate value for II-VI and III-V semiconductors, we now look at the whole
band-structure (not just the fundamental gap) for a few test cases. Here we consider ZnO and GaN for which 
photo-emission data disagree quite remarkably from LSDA calculations. In figure \ref{ZnO.bands} we compare the band 
structure of wurtzite ZnO obtained respectively from LSDA and our ASIC. 

In ZnO, the valence band top (VBT) is predominantly oxygen 2$p$ in character and the conduction band minimum (CBM) is 
essentially zinc 4$s$. With a value of $\sim 0.5$ for the scaling parameter $\alpha$, the ASIC band gap closely matches the experimental gap of E$_g$=3.43~eV, whereas the LDA band-gap is very small ($\sim 0.85$ eV). Some part of the LDA band gap error in ZnO can be traced to an underestimation of the semi-core Zn 3$d$ states. The LDA binding energy for the Zn 3$d$ states is $\sim 5.5 eV$ while photoemission results place them at around $\sim 7.8$ eV. ASIC however rectifies the problem and is in very good agreement with experiment. This results furthermore in the removal of the spurious Zn$_{3d}$-O$_{2p}$ band mixing seen in LDA. An additional feature is that the band-width of the valence band increases considerably as an effect of the downshift of the $d$ manifold. Its worth mentioning that the positions of the Zn 3$d$ levels obtained from ASIC in the case of ZnS, ZnSe, and ZnTe also agree remarkably well with experiment.

\begin{figure}[ht]
\includegraphics[width=0.45\textwidth]{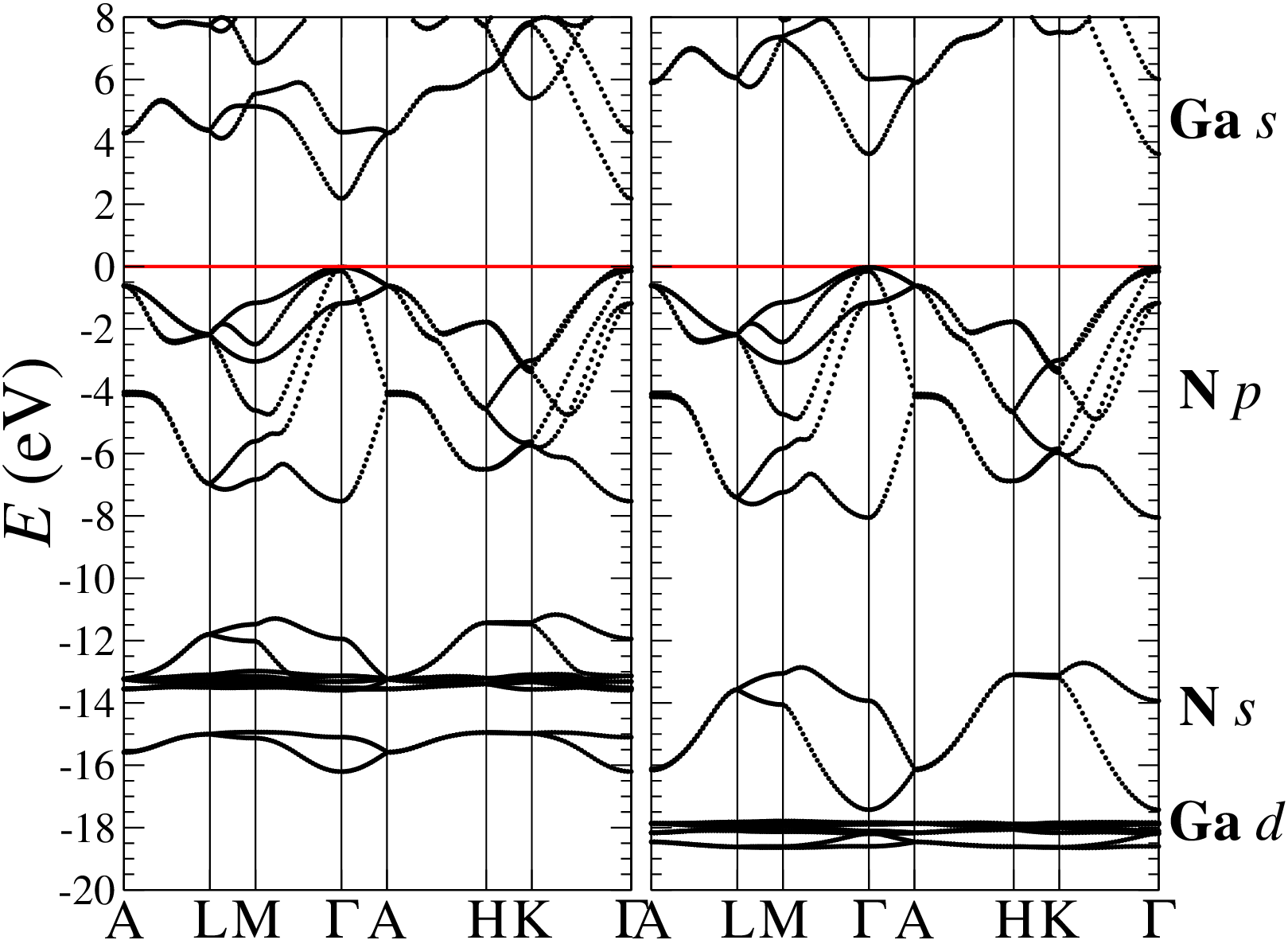}
\caption{\label{GaN.bands}Calculated band structure of wurtzite GaN obtained from LDA and ASIC. The primary orbital character 
of the bands is indicated. The VBT is aligned at 0~eV.}
\end{figure}

The wide-gap III-V semiconductor GaN presents similar phenomenology to that of ZnO. Figure \ref{GaN.bands} compares 
the band structure for wurtzite GaN obtained from LSDA and ASIC. When compared to X-ray photoemission 
spectra \cite{lambrecht}, the LSDA band structure of GaN has several shortcomings. Firstly, the band-gap between N 2$p$ bands 
(VBT) and Ga 4$s$ bands (CBM) is underestimated at around 2.2~eV against the experimental value of 3.4~eV. Secondly, 
the 3$d$ states of Ga are too shallow in LSDA, leading to a spurious 3$d$-2$s$ hybridization.  As a result the Ga 3$d$ states overlap 
with and split the N 2$s$ bands. ASIC rectifies the picture on both counts by improving the band gap and lowering 
the position of the Ga 3$d$ bands with respect to the N 2$s$ bands.
%


\subsection{Transition-metal oxide: MnO}

\begin{figure}[ht]
\includegraphics[width=0.45\textwidth]{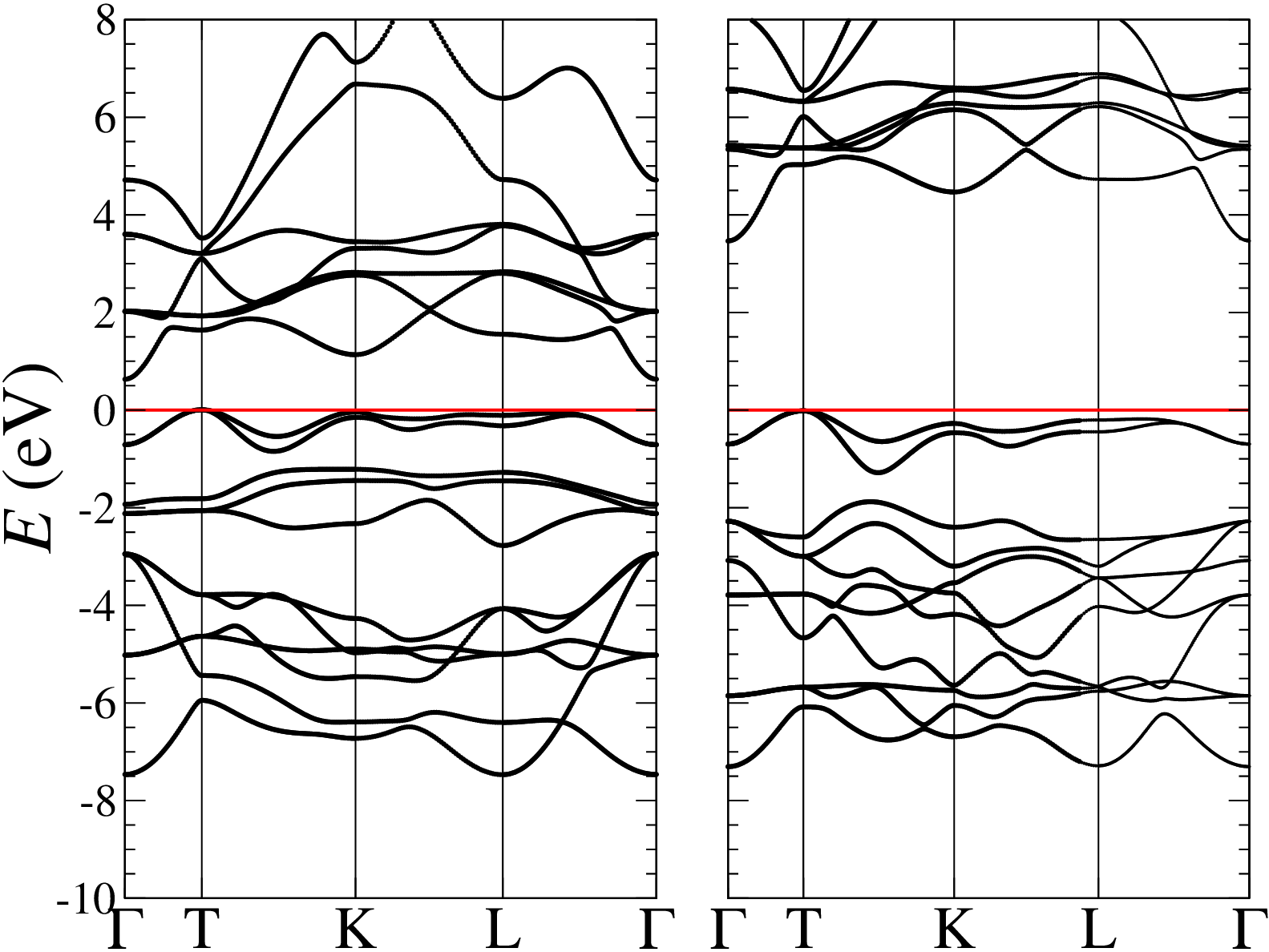}
\caption{\label{MnO.bands}Calculated band structure of anti-ferromagnetic MnO obtained from LSDA and ASIC. 
In our calculation we obtain an LSDA bandgap of $\sim$0.65 eV whereas the ASIC bandgap is much improved 
at $\sim$3.5 eV. The VBT is aligned at 0~eV.}
\end{figure}

Transition metal oxides like MnO and NiO are characterized by partially filled 3$d$ orbitals and an associated local magnetic 
structure. In particular the Mn$^{2+}$ ions in MnO are magnetic with a half-filled 3$d$ shell. In the ground state, MnO is an 
A-type anti-ferromagnetic insulator in the intermediate charge-transfer Mott-Hubbard regime with a band-gap in the region of
3.8-4.2~eV. The VBT is expected to be of mixed Mn 3$d$-O 2$p$ character and the CBM pure Mn 3$d$ in character. However 
the LSDA description of MnO is flawed in several aspects most notably in describing MnO as a narrow gap (E$_g$ = ~0.92 eV) 
Mott-Hubbard insulator with both the VBT and CBM composed of purely of Mn 3$d$ states. This is due to the severe underestimation 
of $d$ electron binding-energies in LSDA. The calculated anti-ferromagnetic band-structures of MnO from LSDA and ASIC ($\alpha=1/2$) are 
shown in figure \ref{MnO.bands}.

Note that these are for the rhombohedral unitcell with 4 atoms per cell. The two Mn ions are anti-ferromagnetically aligned 
and the oxygen ions are non-magnetic. This results in a layered ferromagnetic order of the (111) planes, which in turn
are anti-ferromagnetic coupled to each other. Also in this case, ASIC is a considerable improvement over LSDA. The size of the fundamental gap now resembles the experimental one and the VBT recovers some $p$ character. 


\section{Results: Molecules}

\subsection{Ionization potentials}

In view of the fact that the ASIC method gives improved eigenvalue spectra for several solid state systems, 
it is worth taking a cautious look at how it performs with molecules. This is particularly important for assessing
whether the ASIC scheme can be adapted to work in DFT electron transport schemes based on the KS spectra 
\cite{cormac,smeagol}. In exact KS DFT only the highest occupied orbital eigenvalue $(\epsilon^\mathrm{HOMO})$ 
has a rigorous physical interpretation and corresponds to the negative of the first ionization potential \cite{janak,PPLB}. 
More generally, for a $N$ electron system, the following equations hold in exact KS-DFT
\begin{equation}\label{discharge}
\epsilon^\mathrm{HOMO}(M) = -I_\mathrm{N}~~~~~~~\mathrm{for}~~( N-1 < M < N )
\end{equation}
\begin{equation}\label{charge}
\epsilon^\mathrm{HOMO}(M) = -A_\mathrm{N}~~~~~~~\mathrm{for}~~( N < M < N+1 )
\end{equation}
where $-I_\mathrm{N}$ and $-A_\mathrm{N}$ are the ionization potential (IP) and the electron affinity (EA) respectively. 
Therefore we start our analysis by looking at these quantities as calculated by ASIC. Also in this case we investigate 
different values of $\alpha$. However here we limit ourselves only to $\alpha=1$ (ASIC$_\mathrm{1}$) and 
$\alpha=1/2$ (ASIC$_\mathrm{1/2}$). 

\begin{table}
\caption{\label{IPtab}Experimental Ionization potential (IP) compared to calculated HOMO eigenvalues for neutral 
molecules. Columns 3 and 4 present the results from ASIC with respectively $\alpha=1/2$ and $\alpha=1$. The experimental
data are taken from reference \cite{expmolecules}.}
\begin{ruledtabular}
\begin{tabular}{lcccc}
Molecule&\multicolumn{3}{c}{$\epsilon^\mathrm{HOMO}$(eV)}&-IP(eV)\\
\cline{2-4}\\
&LSDA&ASIC$_\mathrm{1/2}$&ASIC$_\mathrm{1}$&Experiment\\
&    & & &\\
\hline
CH$_3$         	           & -4.65 &  -7.34 &  -10.06 &  -9.84   \\
NH$_{3}$                   & -5.74 &  -8.21 &  -10.79 &  -10.07  \\
SiH$_{4}$                  & -7.95 &  -10.14&  -12.41 &  -11.00  \\
C$_{2}$H$_{4}$             & -6.28 &  -8.00 &  -9.74  &  -10.51  \\
SiCH$_{4}$	           & -5.89 &  -7.57 &  -9.35  &  -9.00   \\
CH$_{3}$CHCl$_{2}$         & -7.23 &  -8.97 &  -10.72 &  -11.04  \\
C$_{4}$H$_{4}$S	           & -5.95 &  -7.65 &  -9.35  &  -8.87   \\
C$_{2}$H$_{6}$S$_{2}$      & -5.56 &  -7.54 &  -9.53  &  -9.30 	 \\
Pyridine                   & -4.83 &  -6.57 &  -8.31  &  -9.60   \\
Benzene	                   & -5.92 &  -7.59 &  -9.28  &  -9.24   \\
Iso-butene                 & -5.39 &  -6.98 &  -8.6   &  -9.22   \\
Nitrobenzene               & -6.49 &  -8.76 &  -10.67 &  -9.92   \\
Naphthalene                 & -5.49 &  -7.04 &  -8.59  &  -8.14   \\
C$_{60}$	           & -5.06 &  -6.53 &  -8.02  &  -7.57 	 \\
C$_{70}$	           & -4.92 &  -6.40 &  -7.89  &  -7.36	 \\
\end{tabular}
\end{ruledtabular}
\end{table}

\begin{figure}[ht]
\includegraphics[width=0.45\textwidth]{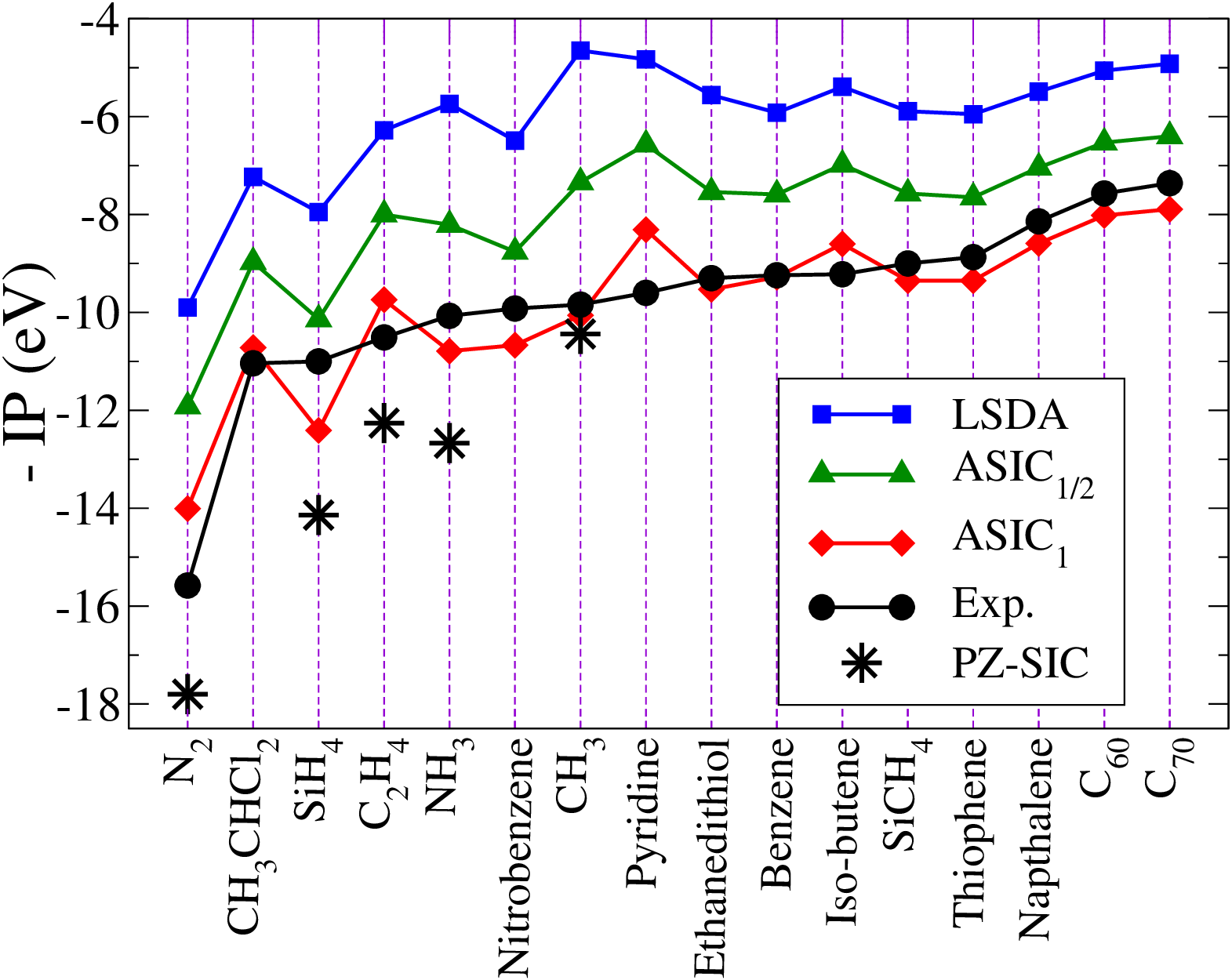}
\caption{\label{IPfig}Experimental negative ionization potential IP compared to the calculated HOMO eigenvalues 
for molecules. The experimental data are from reference \cite{expmolecules}, while the star symbol represents full
PZ-SIC calculations from reference \cite{scuseria2}.}
\end{figure}

In table \ref{IPtab}  and figure \ref{IPfig} we compare the experimental negative IP for several molecules with the corresponding 
$(\epsilon^\mathrm{HOMO})$ obtained using LSDA and ASIC. It is clear that LSDA largely underestimates 
the removal energies in all the cases and that the values obtained from ASIC$_\mathrm{1/2}$ are also consistently 
lower than the experimental value. However, as made evident by the figure the agreement between 
ASIC$_1$ and experiments is surprisingly good. In fact the mean deviation $\delta(X)$ 
($X$ = LSDA, ASIC$_\mathrm{1/2}$, ASIC$_1$) from experiment
\begin{equation*}
\delta(X)=\sqrt{\frac{\sum_{\substack{i=1}}^N \left[\epsilon^\mathrm{HOMO,i}_X+\mathrm{IP}^{i}_\mathrm{Expt}\right]^2}{N}} 
\end{equation*}
is 3.56~eV for LSDA, 1.69~eV for ASIC$_{1/2}$ and only 0.58 eV for ASIC$_1$ ($N$ runs over the molecules of table \ref{IPtab}). 
It is worth noting that we have excellent agreement
over the whole range of molecules investigated going from N$_2$ to large fullerenes C$_{60}$ and C$_{70}$.

For comparison in figure \ref{IPfig} we have also included results obtained with a full self-consistent PZ-SIC 
approach \cite{scuseria2}. Surprisingly our atomic approximation seems to produce a better agreement with experiments
than the self-consistent scheme, which generally overcorrects the energy levels. This is a rather general feature of
the PZ-SIC scheme and it is generally acknowledged that some re-scaling procedure is needed
\cite{vydrov1,vydrov2}. 
%

\begin{table*}[ht]
\caption{\label{EAtab}Calculated HOMO eigenvalues for singly negatively charged molecules compared to experimental 
negative electron affinities (-EA). Columns 6,7 and 8 present the LUMO eigenvalues for the corresponding neutral species.}
\begin{ruledtabular}
\begin{tabular}{lccccccc}
Molecule&\multicolumn{3}{c}{$\epsilon^\mathrm{HOMO}_\mathrm{N+1}$(eV)}&Exp. -EA (eV)
&\multicolumn{3}{c}{$\epsilon^\mathrm{LUMO}_\mathrm{N}$(eV)}\\
\cline{2-4}\cline{6-8}\\
&LSDA& ASIC$_{1/2}$ & ASIC$_1$ &&LSDA& ASIC$_{1/2}$&ASIC$_1$\\
\hline
CN$^-$	            &  0.84     &	-0.79	 &   -2.48    &   -3.86    &	-8.13	&   -9.03    &   -9.42	   \\
C$_2$H$^-$	    &  0.94     &	-0.80	 &   -2.68    &   -2.97    &	-6.91	&   -7.38    &	 -7.48	   \\
CH$_3$S$^-$	    &  2.42     &	 0.65 	 &   -1.14    &   -1.87    &	-5.20	&   -5.31    &	 -5.34	   \\
OH$^-$	            &  3.82     &	 1.09 	 &   -1.80    &   -1.83    &	-0.16	&   -0.43    &	 -0.69	   \\
SiH$_3^-$	    &  4.61     &	 3.13 	 &    1.61    &   -1.41    &	-2.66	&   -3.30    &	 -4.07	   \\
HOO$^-$	            &  3.10     &       -0.07	 &   -3.34    &   -1.08    &	-5.30   &   -6.14    &	 -6.40	   \\
NH$_2^-$	    &  3.83     &	 1.51 	 &   -0.98    &   -0.77    &	-5.27	&   -4.80    &	 -4.39	   \\
CH$_2^-$	    &  3.07     &	 1.21 	 &   -0.45    &   -0.65    &	-3.80	&   -3.84    &	 -3.91	   \\
CH$_3$CO$^-$	    &  2.90     &	 1.76 	 &    0.40    &   -0.42    &	-2.94	&   -3.88    &	 -4.85	   \\
CHO$^-$	            &  3.55     &        2.02    &    0.42    &   -0.31    &    -3.30   &   -4.40    &	 -5.51	   \\
CH$_3^-$	    &  4.15     &	 1.99 	 &   -0.34    &   -0.08    &	-2.73	&   -2.59    &	 -2.47	   \\
C$_{60}^-$	    &  0.03     &	-1.19	 &   -2.45    &   -2.65    &	-3.44	&   -4.66    &	 -5.90	   \\
C$_{70}^-$	    &  0.00     &	-1.22	 &   -2.47    &   -2.73    &	-3.17	&   -4.41    &	 -5.66	   \\
\end{tabular}
\end{ruledtabular}
\end{table*}

\subsection{Electron affinities}

In Hartree Fock theory where Koopmans' theorem holds \cite{koop}, the lowest unoccupied molecular orbital (LUMO) 
energy $(\epsilon^\mathrm{LUMO})$, corresponds to the vertical EA of the $N$ electron system, if one neglects electronic
relaxation. No such interpretation exists for $(\epsilon^\mathrm{LUMO})$ in DFT and so the EA is not directly accessible 
from the ground state spectrum of the $N$ electron system. However, as equation (\ref{charge}) indicates, the EA is in 
principle accessible from the ground state spectrum of the $N+1-f$ ($0 < f < 1$) electron system and asserts in particular 
that it must be relaxation free through non-integer occupation. Unfortunately, the LSDA/GGA approximate functionals usually 
perform rather poorly in this regard as the $N+1$ electron state is unbound with a positive eigenvalue. So one resorts instead 
to extracting electron affinities from more accurate total energy differences \cite{DSCF}, or by extrapolating them from
LSDA calculations for the $N$ electron system \cite{alessioaff}. This failing of approximate 
functionals has been traced in most part to the SI error and so SIC schemes are expected to be more successful in 
describing the $N+1$ electron spectrum.

In table \ref{EAtab} we compare HOMO energies (denoted as $\epsilon^\mathrm{HOMO}_{N+1}$) of several singly 
negatively charged molecules with the experimental electron affinities. We also report the LUMO energies 
for the corresponding neutral species (denoted as $\epsilon^\mathrm{LUMO}_\mathrm{N}$). 
LSDA relaxed geometries for the neutral molecule are used for both the neutral and charged cases. We find that various 
$\epsilon^\mathrm{HOMO}_\mathrm{N+1}$ obtained from ASIC$_1$ once again are in reasonably good 
agreement with corresponding experimental electron affinities while LSDA and ASIC$_{1/2}$ continue to be 
poor even in this regard. In this case $\delta(X)$ stands at 4.1~eV , 2.31~eV and 0.83~eV for LSDA, 
ASIC$_{1/2}$ and ASIC$_1$ respectively. Notice that $\epsilon^\mathrm{HOMO}_\mathrm{N+1}$ from LSDA 
is positive in most cases as the states are unbound. 

\begin{figure}[ht]
\includegraphics[width=0.45\textwidth]{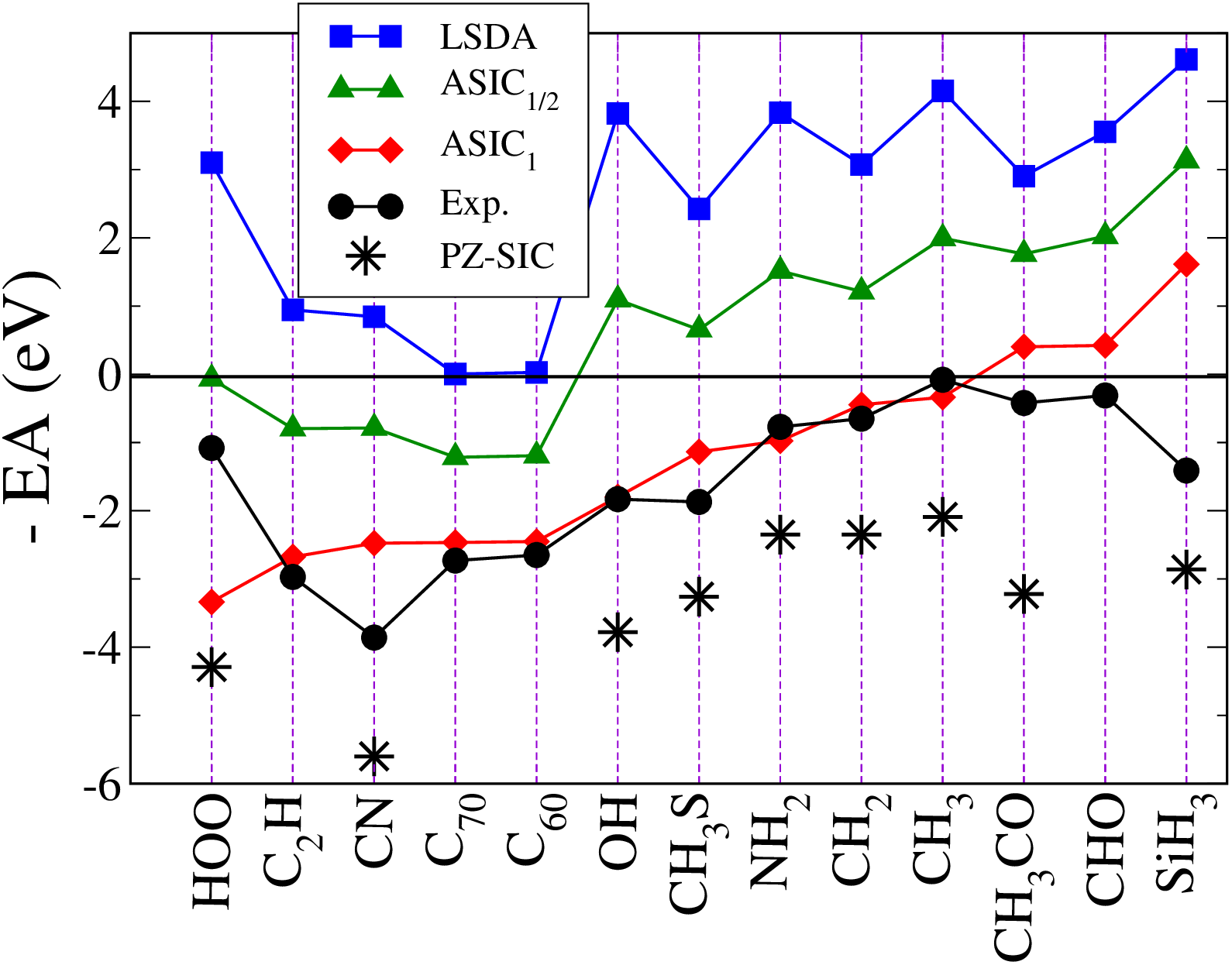}
\caption{\label{EAfig}Experimental negative electron affinities (-EA) compared to calculated HOMO eigenvalues 
of negative radicals.}
\end{figure}

In figure \ref{EAfig} we present our data together with $\epsilon_\mathrm{N+1}^\mathrm{HOMO}$ as calculated 
using the PZ-SIC \cite{scuseria2}. Again ASIC$_1$ performs better than PZ-SIC, that also for the EA systematically
overcorrects.


\subsection{Vertical excitations}

Having shown that ASIC offers a good description of both IP and EA for a broad range of molecules, we turn our
attention to the remaining vertical ionization potentials. As mentioned before, KS-DFT lacks of Koopmans theorem,
and therefore the KS energies are not expected to be close to the negative of the vertical ionization potentials.
However, at least for atoms, the introduction of SIC brings a remarkable cancellation between the negative relaxation energy and
the positive non-Koopmans corrections \cite{PZSIC}. For this reason the SIC KS eigenvalues are a good approximation
to the relaxed excitation energies. As an example, in table \ref{N2tab} we present the orbital energies calculated with ASIC$_1$ and ASIC$_{1/2}$ for the N$_2$ molecule. These are compared with experimental data \cite{N2} and orbital energies obtained
respectively with Hartree-Fock (HF), self-consistent SIC, and SIC where molecular orbitals are used instead of localized orbitals
(D-SIC) \cite{lin2}.
\begin{table}
\caption{\label{N2tab}Orbital energies of N$_2$ calculated with various methods. The results for Hartree-Fock and 
SIC are from reference \cite{lin2}. Experimental results are from reference \cite{N2}.}
\begin{ruledtabular}
\begin{tabular}{lccccccc}
Orbital& HF &SIC&D-SIC&ASIC$_\mathrm{1}$&ASIC$_\mathrm{1/2}$&LSDA&Exp.\\
\hline
2$\sigma_g$&  -41.49  & -38.86 & -37.85 & -38.29 & -33.22 & -28.16 &  \\
2$\sigma_u$& -21.09   & -20.27 & -16.44 & -18.42 & -15.64 & -12.93 & -18.75 \\
3$\sigma_g$&  -17.17  & -17.39 & -13.88 & -14.01 & -11.70 & -9.90 & -15.58 \\
1$\pi_u$       &  -16.98  & -16.33 & -16.68 & -15.97 & -13.74 & -11.54 & -16.93 \\
\end{tabular}
\end{ruledtabular}
\end{table}

Remarkably ASIC$_1$ seems to offer good agreement over the whole spectrum, improving considerably over
LSDA and in some cases even over SIC and HF results. This improvement is not just quantitative, but also qualitative. 
For instance while rectifying the LSDA spectrum of the N$_2$ molecule, ASIC$_1$ preserves the correct order 
between $3\sigma_g$ and $1\pi_u$ orbitals, which are erroneously inverted by both SIC and HF. So why does 
ASIC perform better than the other methods with regards to removal energies? In LSDA, electron relaxation 
typically cancels only half of the non-Koopmans
contributions, resulting in energies that are too shallow \cite{PZSIC}. In contrast HF lacks energy relaxation
and the orbital energies are too deep. The reason why ASIC$_1$ performs better than self-consistent SIC 
is less clear. As a general consideration, also for the case of vertical ionization energies self-consistent SIC 
seems to overcorrect the actual values. Thus the SIC potential appears too deep, and the averaging procedure
behind the ASIC approximation is likely to make it more shallow. 

As a further test we calculated the orbital energies for a few other molecules and compared 
them both with LSDA and experiment \cite{chong}. These are presented in table \ref{VIPtab}.
\begin{table}
\caption{\label{VIPtab}Orbital energies for CO, HF and H$_2$O calculated with LSDA and ASIC$_1$. 
The experimental results are from reference \cite{chong} and references therein.}
\begin{ruledtabular}
\begin{tabular}{llccc}
Molecule~~~& Orbital~~~ &~LSDA~&~ASIC$_\mathrm{1}~$&~Exp.\\
\hline
CO& 5$\sigma$&  -8.74  & -12.85 & -14.01   \\
      &1$\pi$&  -11.54  & -16.64 & -16.91   \\
      & 4$\sigma$&  -13.97  & -19.36 & -19.72   \\
HF &1$\pi$&  -9.83  & -16.96 & -16.19   \\
      & 3$\sigma$&  -13.61  & -19.68 & -19.90   \\
H$_2$O& 1b$_1$&  -7.32  & -13.38 & -12.62   \\
               &3a$_1$&  -9.32  & -14.66 & -14.74   \\
               & 1b$_2$&  -13.33  & -18.03 & -18.55   \\
\end{tabular}
\end{ruledtabular}
\end{table}
Again the ASIC$_1$ results compare rather well with experiment, and we can conclude that the ASIC method
offers a rather efficient and inexpensive theory for single particle vertical excitations. 


\subsection{HOMO-LUMO gap and discontinuity of the exchange and correlation potential}

We are now in a position to discuss the HOMO-LUMO gap in ASIC. As already mentioned,
even for the exact XC functional, the KS gap $E_g^\mathrm{KS}=
\epsilon^\mathrm{LUMO}-\epsilon^\mathrm{HOMO}$ does not account for the actual quasi-particle gap $E_g=I_\mathrm{N}-A_\mathrm{N}$. 
This in turn is the sum of $E_g^\mathrm{KS}$ and the discontinuity of the exchange and correlation potential $\Delta_\mathrm{xc}$.
Equivalently
\begin{equation}
\Delta_\mathrm{xc} = \lim_{\substack{f\rightarrow0^+}}\epsilon^\mathrm{HOMO}_{N+f}-\epsilon^\mathrm{LUMO}_N\:,
\end{equation}
i.e. $\Delta_\mathrm{xc}$ is the discontinuity in the eigenvalue of the LUMO state at $N$. Therefore, in order to extract the 
actual gap from the KS gap, provided that the spectrum is reasonably well described at integer electron numbers $N$, 
what remains is to model the derivative discontinuity at $N$ and ensure that $\epsilon^\mathrm{HOMO}_{N\pm f}$ is 
relaxation free for ($0<f<1$). Local and semi-local (LSDA/GGA) XC functionals lack such a discontinuity, while 
self-interaction corrections are able to restore it, at least in part. For instance the PZ-SIC scheme is successful in this regard.
\begin{figure}[ht]
\includegraphics[width=0.45\textwidth]{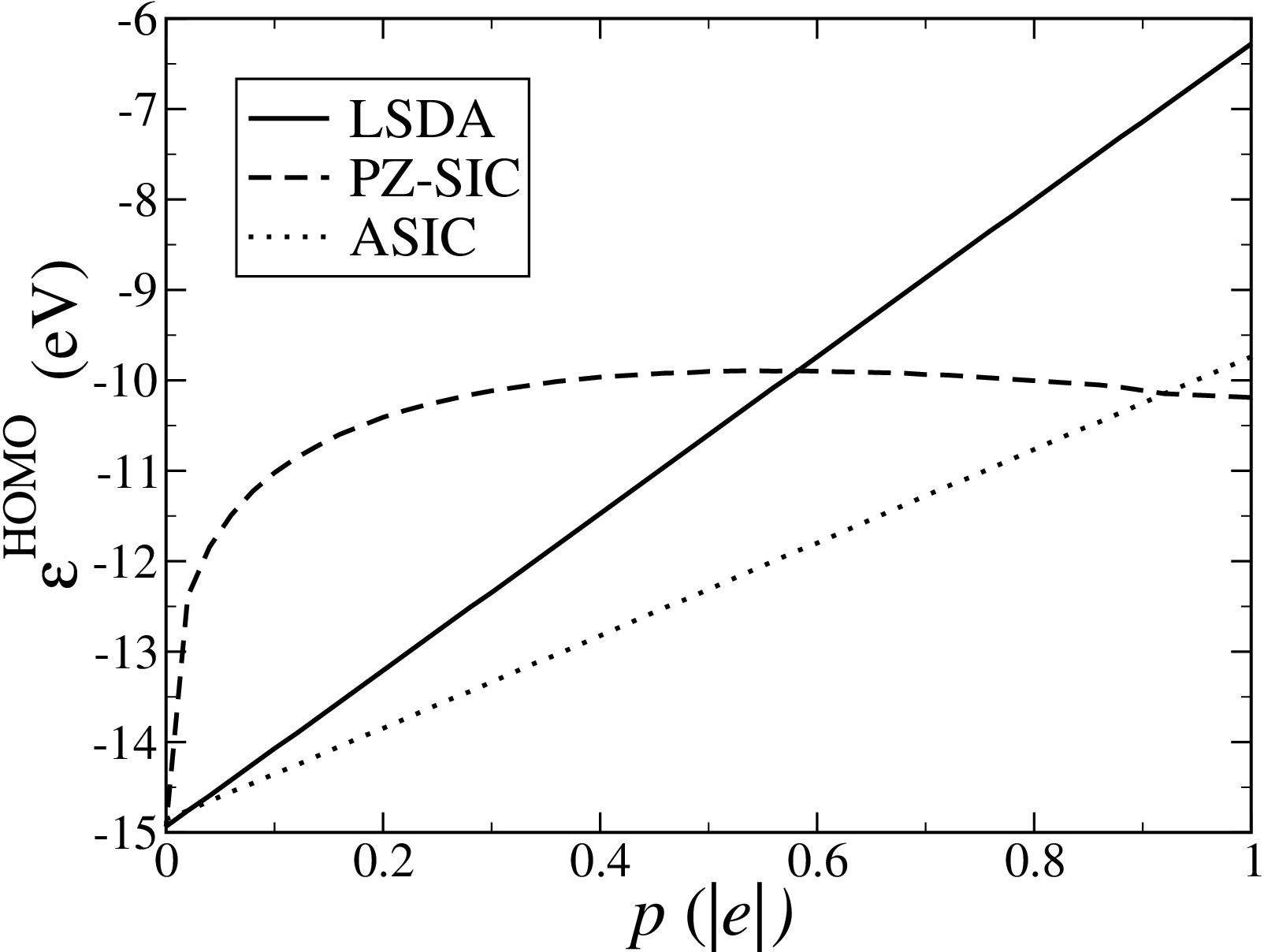}
\caption{\label{Ethylene}Ionization curve for the ethylene (C$_2$H$_4$) molecule as the occupation of the HOMO state 
is varied from 0 to 1 in going from the ionized C$_2$H$_4^+$ to the netural C$_2$H$_4$ state.}
\end{figure}

In figure \ref{Ethylene} we illustrate the ionization curve for the ethylene (C$_2$H$_4$) molecule as the occupation 
of the HOMO state is varied from 0 to 1 in going from the ionized C$_2$H$_4^+$ to the netural C$_2$H$_4$ configuration. 
It is seen that among the three schemes presented, only the PZ-SIC scheme approximately models the behaviour required 
by the equation (\ref{discharge}). The ASIC HOMO eigenvalue roughly agrees with the PZ-SIC eigenvalue at integer occupation 
but behaves linearly through non-integer values. Thus we find that the derivative discontinuity for the molecule is smoothed 
out in ASIC, which still connects continuously different integer occupations. This is one of the limitations of the atomic 
representation employed in ASIC. 

In view of the foregoing discussion, the actual size of the HOMO-LUMO gap in ASIC becomes significant with a direct bearing 
on the physics described. Ideally, we want $\epsilon^\mathrm{LUMO}(N)$ (LUMO for the $N$-electron system) to be as close 
to $\epsilon^\mathrm{HOMO}(N+1)$ so that the range of eigenvalue relaxation through fractional occupation numbers 
$M\in(N,N+1)$ is minimized. Looking at columns 6,7 and 8 in table \ref{EAtab} however, we see that for almost all the 
molecules, this is hardly the case. The agreement between $\epsilon^\mathrm{LUMO}(N)$ and -EA from experiment ($\simeq 
\epsilon^\mathrm{HOMO}(N+1)$) for ASIC$_1$ is quite poor implying a considerable energy range spanning fractional 
particle number. We still expect this energy range to be smaller for ASIC$_1$ than LSDA. It is also apparent from the table 
\ref{EAtab} that $\epsilon^\mathrm{LUMO}_\mathrm{ASIC}(N)$ usually differs from $\epsilon^\mathrm{LUMO}_\mathrm{LSDA}(N)$ 
and in fact by considerable magnitudes in some cases.  Thus the ASIC$_1$  ``correction" to the empty LUMO state does not 
vanish in contrast to the PZ-SIC scheme where, by definition, the empty eigenstates are SIC free. 

Since the SIC operator ${v}_\mathrm{ASIC}^\sigma$ is constructed in an atomic orbital representation, the correction to any KS 
eigenstate $\psi_{n}^\sigma$ either filled or empty
\begin{equation}\label{sicexpv}
\delta E^{n\sigma}_\mathrm{ASIC} = \langle\psi_{n}^\sigma|{v}_\mathrm{ASIC}^\sigma|\psi_{n}^\sigma\rangle
\end{equation}
is not necessarily zero unless $\psi_{n}^\sigma$ only projects onto empty atomic orbitals.
Also this correction to the LUMO with respect to the LSDA is negative in most cases, exceptions being NH$_2$ and 
CH$_3$ where it is desirably positive. Thus the fundamental HOMO-LUMO gap in ASIC is a combination of both the 
HOMO and LUMO corrections. Table \ref{HOLUtab} shows how this combination works out in ASIC$_{1/2}$ and 
ASIC$_1$ when compared to LSDA. The molecular test set is the same as that in table \ref{IPtab}.
\begin{table}
\caption{\label{HOLUtab} HOMO-LUMO gap obtained from ASIC compared to the LSDA value. The values
marked with $*$ correspond to unbound LUMO levels.}
\begin{ruledtabular}
\begin{tabular}{lccc}
Molecule&\multicolumn{3}{c}{$\epsilon^\mathrm{LUMO}-\epsilon^\mathrm{HOMO}$ (eV)}\\
\cline{2-4}\\
&LSDA&ASIC$_{1/2}$&ASIC$_1$\\
\hline
CH$_3$         	           &	1.92	    &  4.75  &      7.59	 \\
NH$_{3}$                   &	7.1$^*$	    &  9.29$^*$   &      11.61$^*$	 \\
SiH$_{4}$                  &	8.44$^*$	    &  9.68   &      10.94	 \\
C$_{2}$H$_{4}$             &	5.81	    &  6.59   &      7.38	 \\
SiCH$_{4}$	           &  	6.19        &  7.07   &      8.06        \\
CH$_{3}$CHCl$_{2}$         &    5.79	    &  6.84   &      7.88	 \\
C$_{4}$H$_{4}$S            &    4.46	    &  5.13   &	     5.8	 \\
C$_{2}$H$_{6}$S$_{2}$      &    4.44	    &  6.02   &	     7.6	 \\
Pyridine    		   &    3.85	    &  4.56   &	     5.26        \\
Benzene	    	           &    5.22	    &  5.9    &	     6.59        \\
Iso-butene  	           &    4.88	    &  5.56   &	     6.26        \\
Nitrobenzene	           &    3.25	    &  4.03   &	     4.42        \\
Naphthalene  	           &    3.36	    &  3.83   &	     4.29        \\
C$_{60}$    	           &    1.62	    &  1.87   &	     2.12        \\
C$_{70}$                   &    1.75        &  1.99   &      2.23        \\
\end{tabular}
\end{ruledtabular}
\end{table}

We see in almost all cases the ASIC gap is systematically larger than the LSDA one. This is expected because 
the correction to the HOMO is usually much stronger than that to the LUMO. In general, ASIC is expected to work 
well for systems where the occupied and un-occupied KS eigenstates of the extended system have markedly 
different atomic orbital signatures being derived predominantly from filled and empty atomic orbitals respectively. 
In such a case, the ASIC correction to the empty states would be nullified in being scaled by near-zero 
atomic orbital populations. In some cases, provided phase factors combine suitably, the correction to the empty states 
can even be positive with respect to the same in LSDA.\\


\subsection{Final Remarks}

Before we conclude, we discuss some general properties of the ASIC method which are relevant to any 
orbital dependent SIC implementation and also some possible pitfalls.
As with other SIC schemes, ASIC is not invariant under unitary 
transformations of the orbitals used in constructing the SIC potential. Thus the ASIC correction is likely 
to change as the atomic orbitals used for projecting onto the KS eigenstates of the system are rotated or 
transformed otherwise. Unlike the Perdew-Zunger method however, there can be no variational principle 
over all possible unitary transformations of the atomic orbitals because in the general case they do not 
represent the Hamiltonian of the system under consideration. This also implies that if the scheme is used 
with a system that is already well described by LSDA, the ``correction" additional to the LSDA result does 
not necessarily vanish. Simple metals and narrow gap systems are likely candidates for this scenario.

Furthermore, its pertinent to mention that ASIC becomes ineffective if not counterproductive for materials
with homonuclear bonding, in which valence and conduction bands have the same atomic orbital
character. In this situation the ASIC potential will shift the bands in an almost identical way, without 
producing any quantitative changes, such as the opening up of the KS gap. Note that this is a pitfall of the ASIC
approximation, which distinguishes occupied from empty states only through their projected atomic orbital occupation,
but not of the SIC in general. Typical cases are those of Si and Ge. The KS gap in Si goes from 0.48~eV in
LSDA to only 0.57~eV for ASIC$_{1/2}$, while Ge is a metal in both cases. In addition the LSDA calculated 
valence bandwidths of 12.2~eV for Si and 12.8~eV for Ge, in good agreement with experiments, are 
erroneously broadened to 14.3~eV and 14.8~eV respectively.

\section{Conclusions}

In conclusion, we have implemented the ASIC scheme proposed by Filipetti and Spaldin within the 
pseudopotential and localized orbital framework of the {\it Siesta} code. We have then investigated a
broad range of semiconductors and molecules, with the aim of providing a reasonable estimate for the
scaling parameter $\alpha$. We found that $\alpha=1$, which accounts for the full atomic SI, describes
surprisingly well ionic semiconductors and molecules. In particular for molecules, both the IP and the EA
can be obtained with good accuracy from the HOMO KS eigenvalues respectively for the neutral and 
singly charged molecule. This makes the ASIC scheme particularly suited for application such as quantum 
transport, where the position of the HOMO level determines most of the $I$-$V$ curve. 

In contrast III-V and II-VI semiconductors are better described by $\alpha=1/2$, which corrects
the atomic SI for screening. This makes ASIC$_{1/2}$ an interesting effective band theory for semiconductors. 
The relation of the present scheme with the fully self-consistent SIC methods has been emphasized, and so
has been that with LDA+$U$.

\section{Acknowledgments}
We acknowledge stimulating discussions with Malgorzata Wierzbowska, Kieron Burke and Alessio Filippetti.
This work is supported by the Science Foundation of Ireland under the grants SFI02/IN1/I175.
DSP acknowledges financial support from the University of the Basque Country (Grant No. 
9/UPV 00206.215-13639/2001), the Spanish Ministerio de Educacion y Ciencia 
(Grant No. FIS2004-06490-C03-00) and the Basque Goverment and Diputacion Foral de
Giupuzkoa (Grant NANOMATERIALES and NANOTRON, ETORTEK program).
Traveling has been sponsored by Enterprise Ireland under the International Collaboration 
programme EI-IC/2003/47. Computational resources have been provided by the HEA IITAC 
project managed by the Trinity Center for High Performance Computing and by
ICHEC.

\end{document}